\documentclass[12pt]{article}
\usepackage{graphics}
\usepackage{amssymb}
\usepackage{psfrag}

\newcommand{\rf}[1]{(\ref{#1})}
\newcommand{\beq}{\begin{equation}}
\newcommand{\eeq}{\end{equation}}
\newcommand{\bea}{\begin{eqnarray}}
\newcommand{\eea}{\end{eqnarray}}

\newcommand{\e}{\mbox{e}}

\renewcommand{\a}{\alpha}

\newcommand{\del}{\delta}
\newcommand{\Del}{\Delta}
\newcommand{\sg}{\sigma}

\newcommand{\prt}{\partial}
\newcommand{\mi}{\!-\!}
\newcommand{\equ}{\!=\!}
\newcommand{\pl}{\!+\!}

\newcommand{\cD}{{\cal D}}

\begin{document}

{\normalsize \hfill SPIN-05/14}\\
\vspace{-1.5cm}
{\normalsize \hfill ITP-UU-05/18}\\
${}$\\

\begin{center}
\vspace{48pt}
{ \Large \bf  Reconstructing the Universe}

\vspace{40pt}

{\sl J. Ambj\o rn}$\,^{a,c}$,
{\sl J. Jurkiewicz}$\,^{b}$
and {\sl R. Loll}$\,^{c}$

\vspace{24pt}
{\footnotesize

$^a$~The Niels Bohr Institute, Copenhagen University\\
Blegdamsvej 17, DK-2100 Copenhagen \O , Denmark.\\
{ email: ambjorn@nbi.dk}\\

\vspace{10pt}

$^b$~Mark Kac Complex Systems Research Centre,\\
Marian Smoluchowski Institute of Physics, Jagellonian University,\\
Reymonta 4, PL 30-059 Krakow, Poland.\\
{email: jurkiewicz@th.if.uj.edu.pl}\\

\vspace{10pt}

$^c$~Institute for Theoretical Physics, Utrecht University, \\
Leuvenlaan 4, NL-3584 CE Utrecht, The Netherlands.\\
{email:  j.ambjorn@phys.uu.nl, r.loll@phys.uu.nl}\\

\vspace{10pt}

}
\vspace{48pt}

\end{center}


\begin{center}
{\bf Abstract}
\end{center}

\noindent
We provide detailed evidence for the claim that nonperturbative
quantum gravity, defined through state sums of causal
triangulated geometries, possesses a large-scale limit in which
the dimension of spacetime is four and the dynamics of the
volume of the universe behaves semiclassically. This is a
first step in reconstructing the universe from a
dynamical principle at the Planck scale, and at the same time
provides a nontrivial consistency check of the method of
causal dynamical triangulations.
A closer look at the quantum geometry reveals
a number of highly nonclassical aspects, including a dynamical
reduction of spacetime to two dimensions on short scales and a fractal 
structure of slices of constant time.

\vspace{12pt}
\noindent

\newpage

\section{Introduction}\label{intro}

Nonperturbative quantum gravity can be defined as the quest for
uncovering the true dynamical degrees of freedom of spacetime
geometry at the very shortest scales. Because of the enormous
quantum fluctuations predicted by the
uncertainty relations, geometry near the Planck scale will be
extremely rugged and
nonclassical. Although different approaches
to quantizing gravity do not agree on the precise
nature of these fundamental excitations, or on how they can be determined,
most of the popular formulations\footnote{An exception are
renormalization group approaches which look for nontrivial
ultraviolet fixed points in a metric formulation \cite{reuter,lauscher,reuteretc}.} 
agree that they are neither the smooth 
metrics $g_{\mu\nu}(x)$ (or equivalent classical field variables) of 
general relativity nor straightforward quantum analogues
thereof. In such scenarios, one expects the metric to re-emerge
as an appropriate description of spacetime geometry 
only at larger scales. 

Giving up the spacetime metric at the Planck scale does not 
mean discarding geometry
altogether, since geometric properties such as the presence of a distance
function pertain to much more general structures than differential manifolds
with smooth metric assignments. One could hope
that quantum gravity, when formulated in terms of the conjectured new
microscopic degrees of freedom, was better behaved in the ultraviolet
regime than standard perturbative approaches based on the spacetime
metric. At the same time, it
becomes a nontrivial test for such nonperturbative theories of
quantum gravity whether they can reproduce the correct classical limit
at sufficiently large scales. However,
using this as a consistency check to discriminate between good and bad
candidate theories is in practice complicated by the fact that {\it some}
explicit results about the quantum dynamics of the
proposed quantum gravity theory must be known. For example,
this is not yet the case in loop quantum gravity \cite{thie,rovelli,npz}
or in four-dimensional
spin foam models for gravity \cite{perez}.

In the method of {\it Causal Dynamical Triangulations}
one tries to construct a theory of quantum gravity as
a suitable continuum limit of a superposition of spacetime
geometries \cite{al,ajl1,ajl4d}. In close analogy with
Feynman's famous path integral for the nonrelativistic particle,
one works with an intermediate regularization in which the
geometries are piecewise flat\footnote{These are the analogues of the
piecewise straight paths of Feynman's approach. However, note that
the geometric configurations of the quantum-gravitational path integral
are {\it not} imbedded into a higher-dimensional space, and therefore
their geometric properties such as piecewise flatness are intrinsic,
unlike in the case of the particle paths.}. The primary 
object of interest in this approach
is the propagator between two boundary configurations
(in the form of an initial and final spatial geometry), which
contains the complete dynamical information about the quantum
theory. Because of the calculational complexity of the full,
nonperturbative sum over geometries (the ``path integral"), an
analytical evaluation is at this stage out of reach. Nevertheless,
powerful computational tools, developed in Euclidean quantum
gravity \cite{aj,aj1,bielefeld,bielefeld2,simon,simon2,smit} and other
theories of random geometry (see \cite{book}
for a review), can be brought to bear on the problem.

This paper describes in detail how Monte Carlo simulations have
been used to extract information about the quantum theory, and
in particular, the geometry of the quantum ground state\footnote{Here
and in the following, by ``ground state" we will always mean the state 
selected by Monte Carlo simulations, performed under the constraint 
that the volume of spacetime is (approximately) kept fixed, a
constraint we have to impose for simulation-technical reasons.} 
dynamically
generated by superposing causal triangulations. It follows the
announcement of several key results in this approach to quantum
gravity, first, a ``quantum derivation" of the fact that spacetime is
macroscopically four-dimensional \cite{ajl-prl}, second, a
demonstration that
the large-scale dynamics of the spatial volume of the universe
(the so-called ``scale factor") observed in causal dynamical
triangulations can be described by an effective action closely
related to standard quantum cosmology \cite{semi}, 
and third, the discovery that in the limit of short
distances, spacetime becomes effectively two-dimensional,
indicating the presence of a dynamically generated ultraviolet 
cutoff \cite{newpaper}. 
They do not provide conclusive proof that our construction
from first principles
{\it does} lead to a viable theory of quantum gravity,
but go some way in establishing the existence of a
physically meaningful classical limit. In addition, our
detailed geometric analysis throws some light on the highly
nonclassical microscopic properties of the spacetimes
dominating the gravitational path integral, which after all
are higher-dimensional analogues of the nowhere-differentiable
paths that provide the support of Feynman's path integral.

Our presentation will proceed as follows. After recollecting some
basic properties of the geometric set-up of the regularized path
integral over causal triangulations in Sec.\ \ref{setup}, 
we explain in Sec.\ \ref{num}
how the path integral is implemented numerically. The statistical
model underlying our construction possesses three phases,
which are distinguished by the geometric nature of their ground
states, as described in Sec.\ \ref{phase}. From this point on, 
we concentrate on the phase which yields a ground state
which is extended in time and space. In Sec.\ \ref{measure}
we provide the detailed numerical evidence for why this
state is four-dimensional at large scales and why its shape can be
described by a minisuperspace action. 
We then embark on
a detailed investigation of substructures of the spacetime
universe, namely, spatial slices (of fixed integer time $\tau$) in Sec.\ \ref{thin} 
and in Sec.\ \ref{thick}
``thick spatial slices", which are geometries of finite time
extension $\Delta \tau\equ 1$. We measure both their Hausdorff
and spectral dimensions, the latter from a discrete
diffusion process. Interestingly, the geometry of the ``thin" spatial slices
shares some characteristics with a class of branched
polymers which has already appeared in areas as diverse as
noncritical string theory, econophysics and network theory.
A summary of our results and an outlook are 
presented in Sec.\ \ref{conclusion}.

\section{Geometric set-up and simplicial action}\label{setup}

\begin{figure}[ht]
\centerline{\scalebox{0.6}{\rotatebox{0}{\includegraphics{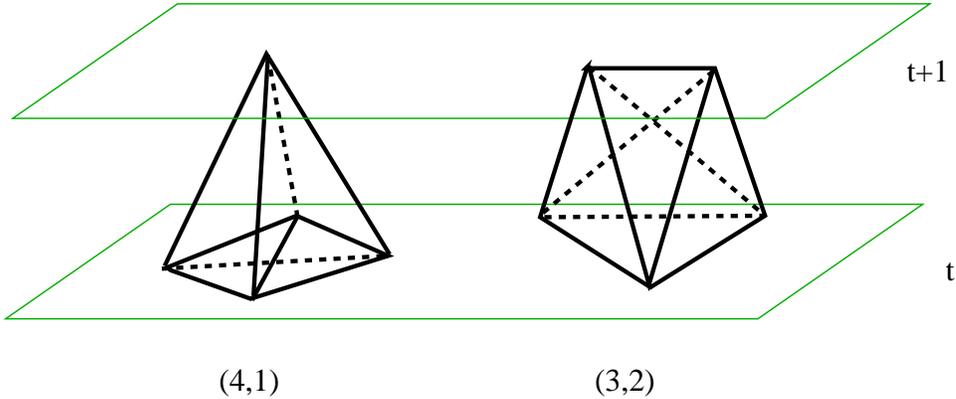}}}}
\caption[phased]{{\small
The two fundamental building blocks of causal dynamically triangulated gravity.
The flat four-simplex of type (4,1) on the left has four of its vertices at time $\tau$ and
one at time $\tau\pl 1$, and analogously for the (3,2)-simplex on the right. The ``gap"
between two consecutive spatial slices of constant integer time is filled by copies
of these simplicial building blocks and their time-reversed counterparts,
the (1,4)- and the (2,3)-simplices.
}}
\label{4dsimplices}
\end{figure}

We briefly remind the reader of the structure of the
spacetime geometries which contribute to the regularized
path integral \cite{ajl4d}. The fundamental building blocks
are two types of four-simplices ``cut out of" four-dimensional flat
Minkowski space. A flat four-simplex of type (4,1) has six space-like
edges of length-squared $l_{\rm space}^2\equ a^2$ and four time-like
edges of length-squared $l_{\rm time}^2\equ -\alpha a^2$. Similarly, a flat
four-simplex of type (3,2) has four space-like edges of length-squared
$l_{\rm space}^2\equ a^2$ and six time-like ones of length-squared
$l_{\rm time}^2\equ -\alpha a^2$. Each spacetime has a well-defined
global foliation by a ``proper time" $\tau$.\footnote{We choose the name
``proper time" because the a priori integer-valued variable $\tau$ can
be extended continuously to the entire spacetime \cite{blackhole}. 
Within each simplex
one can choose a coordinate system whose time variable coincides
with $\tau$ and also with the proper time of a freely falling observer in
the simplex. Note that our setting here is that of piecewise flat and not of smooth
geometries.} Fig.\ \ref{4dsimplices} illustrates how the
simplicial building blocks are situated between consecutive spatial
slices of constant integer $\tau$. The (4,1)-simplex consists of a spatial
tetrahedron at time $\tau$ whose four vertices are connected by four
time-like edges to the single vertex at time $\tau\pl 1$. Similarly, the
(3,2)-simplex consists of a spatial triangle at time $\tau$ whose vertices
are connected to the two vertices at $\tau\pl 1$ which form the end points
of the fourth spatial edge of the simplex.

The way in which these building blocks, together with their counterparts with
inversed time orientation, can form a spacetime slice of height $\Delta \tau\equ 1$ is
by gluing them pairwise along a shared time-like
boundary tetrahedron. In order to get a spacetime of time duration $\tau\equ t$,
one glues together
$t$ such slices in succession, which is of course only possible whenever
their spatial boundary geometries at integer-$\tau$ match pairwise.
We require the gluings to be such that the resulting
discretized spacetime is a simplicial manifold. The
spacetime topology is fixed to $[0,1]\times {}^{(3)}\Sigma$, where for our
investigations we have chosen as constant-$\tau$ surfaces the
three-spheres $ {}^{(3)}\Sigma \equ S^3$.
For convenience in the computer simulations, we also sometimes identify
time periodically, leading to a spacetime topology $S^1\times S^3$.

From the edge length assignments one can compute volumes and interior
angles of the simplices which are needed because they appear in the
Regge form of the Einstein action, from which the weight of each triangulated
spacetime $T$
in the sum over geometries will be computed (see \cite{ajl4d} for calculational
details). The gravitational action becomes
then a simple function of a number of positive integers, which count the
overall numbers of simplices or subsimplices of a certain type.
For completeness, we give the explicit form of the action
after analytic continuation to the Euclidean sector, as a function of
the number
$N_0$ of vertices (zero-simplices), and the numbers $N_4^{(3,2)}$ and
$N_4^{(4,1)}$ of four-simplices of types (3,2) and (4,1) of either
time orientation,
\begin{eqnarray}
&&\!\!\! S_E=-k^{(b)}\pi \sqrt{4\tilde\alpha -1}(N_0-\chi)\nonumber\\
&&+ N_{4}^{(4,1)} \Biggl( k^{(b)}\sqrt{4\tilde\alpha -1} \biggl(  -\frac{\pi}{2}
-\frac{\sqrt{3}}{\sqrt{4\tilde\alpha -1}}
 \arcsin\frac{1}{2\sqrt{2}\sqrt{3\tilde\alpha -1}} \nonumber\\
&&\;\;\;\;\;\;\;\;\;\;\;\;\;\;\; +\frac{3}{2}
\arccos\frac{2\tilde\alpha -1}{6\tilde \alpha -2}\biggr)+
\lambda^{(b)} \frac{\sqrt{8\tilde\alpha -3}}{96} \Biggr)\nonumber\\
&&+N_{4}^{(3,2)} \Biggl( k^{(b)} \sqrt{4\tilde\alpha -1} \biggl( -\pi
+\frac{\sqrt{3}}{4\sqrt{4\tilde\alpha -1}}\arccos
\frac{6\tilde\alpha -5}{6\tilde \alpha -2}
+\frac{3}{4} \arccos\frac{4\tilde\alpha -3}{8\tilde \alpha -4}
\nonumber\\
&&\;\;\;\;\;\;\;\;\;\;\;\;\;\;\; +\frac{3}{2}
\arccos\frac{1}{2\sqrt{2}\sqrt{2\tilde\alpha -1}\sqrt{3\tilde\alpha -1}}\biggr)
+\lambda^{(b)} \frac{\sqrt{12\tilde\alpha -7}}{96} \Biggr).
\label{act4dis}
\end{eqnarray}
In this expression, $\chi$ is the Euler characteristic of the piecewise flat
four-manifold and $\tilde\alpha \equiv -\alpha$ denotes the positive
ratio between the two types of squared edge lengths after the
Euclideanization. In order to satisfy the triangle inequalities, we need
$\tilde\alpha >7/12$ \cite{ajl4d}. For simplicity,
we have assumed that the manifold
is compact without boundaries. In the presence of boundaries, appropriate
boundary terms must be added to the action.

For the simulations, a convenient alternative parametrization of the action
is given by
\begin{equation}
S_E=-(\kappa_0+6\Delta) N_0+\kappa_4 (N_{4}^{(4,1)}+N_{4}^{(3,2)})+
\Delta (2 N_{4}^{(4,1)}+N_{4}^{(3,2)}),
\label{actshort}
\end{equation}
where the functional dependence of the $\kappa_i$ and $\Delta$ on the
bare inverse Newton constant $k^{(b)}$, the bare cosmological constant
$\lambda^{(b)}$ and $\tilde\alpha$ can be computed from \rf{act4dis}. We have
dropped the constant term proportional to $\chi$, because it will be irrelevant
for the quantum dynamics. Note that $\Delta\equ 0$ corresponds to
$\tilde\alpha\equ 1$, and $\Delta$ is therefore a measure of the asymmetry
between the lengths of the spatial and timelike edges of the simplicial
geometry. As we shall
see, the physically interesting region of the phase diagram of four-dimensional
causal dynamical triangulations has positive $\Delta$, which translates into
values of $\tilde\alpha$ smaller than 1.

\section{Numerical implementation}\label{num}

We have investigated the infinite-volume limit of
the ensemble of causal triangulated four-dimensional
geometries with the help of Monte Carlo simulations at finite
four-volumes $N_4\equ N_4^{(4,1)}+N_4^{(3,2)}$ of up to 362.000
four-simplices. A simplicial geometry is stored in the computer as a
set of lists, where the lists consist of dynamic sequences of labels
for simplices of dimension $n$, $0\leq n\leq 4$, together with their
position and orientation with respect to the time direction. Additional
list data
include information about nearest neighbours, i.e. how the
triangulation ``hangs together", and other discrete data
(for example, how many four-simplices meet at a given edge) 
which help improve the acceptance rate of Monte Carlo moves.

The simulation is set up to generate a random walk in
the ensemble of causal geometries of a fixed time extension $t$.
The local updating algorithm consists of a set of moves that
change the geometry of the simplicial
manifold locally, without altering its topological properties. These
can be understood as a Lorentzian variant of (a simplified version of) the so-called
Alexander moves \cite{alexander,pachner,gv}, in the sense that they are compatible
with the discrete time slicing of our causal geometries.
For example, the subdivision of
a four-simplex into five four-simplices by placing a new vertex
at its centre is not allowed, because vertices can only be
located at integer times $\tau$. Details of the local moves can be
found in \cite{ajl4d}.
As usual, each suggested local change of triangulation
is accepted or rejected according to
certain probabilities depending on the change in the action
and the local geometry. (Note that a move will always be rejected if
the resulting triangulation violates the simplicial manifold property.)
The moves are called in random order, with probabilities chosen in
such a way as to ensure
that the numbers of actually performed moves of each type are
approximately equal. We attained a rather high average acceptance
rate of about 12.5\%, which was made possible by keeping track
of sub-simplex structures relevant for the performance
of certain moves, as mentioned earlier. The attempted
moves were performed
in units of $10^6$ (one ``sweep"), with the time for one
sweep approximately independent of the system volume in the
region $\kappa_0\approx 2.2$ and $\Delta\approx 0.4\dots 0.6$ where
most of our measurements were taken.

\begin{figure}[t]
\vspace{-2cm}
\psfrag{V}{\bf{\Large $N_4^{(3,2)}$}}
\psfrag{P(V)}{\Large\bf $P(N_4^{(3,2)})$}
\centerline{\scalebox{0.6}{\rotatebox{-90}{\includegraphics{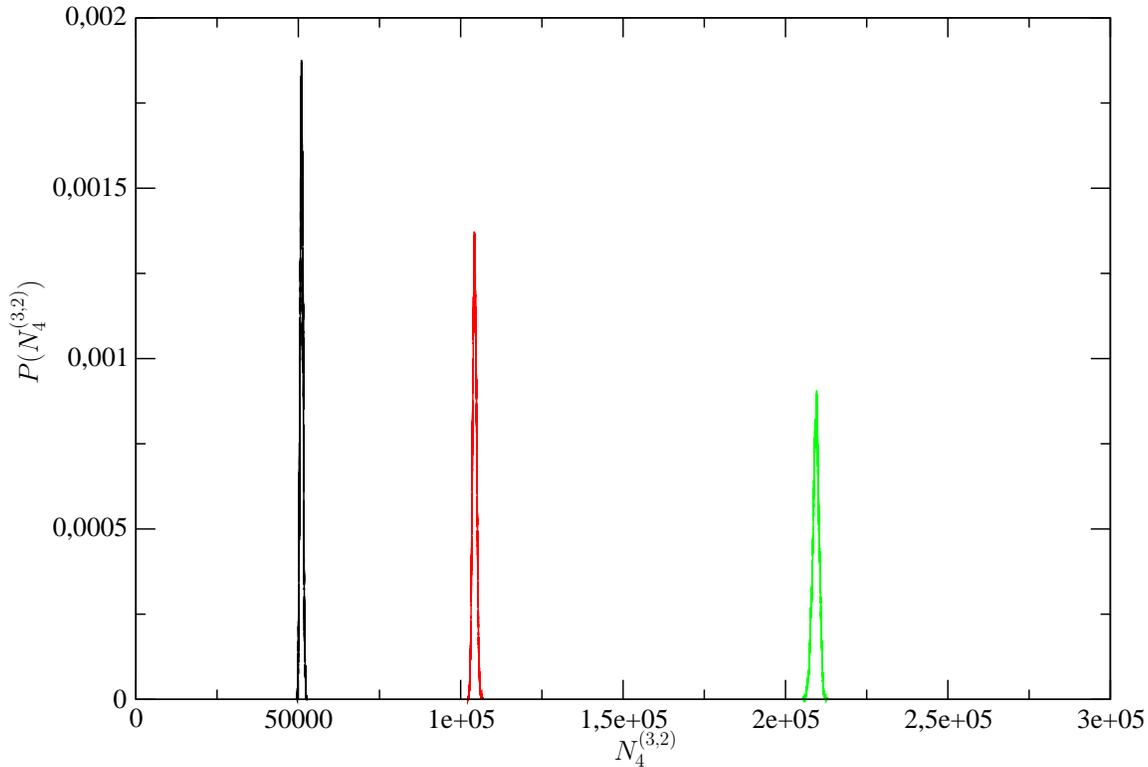}}}}
\vspace{-0.0cm}
\caption[phased]{{\small Unnormalized distribution of the number
$N_4^{(3,2)}$ of four-dimensional
simplices of type (3,2), at fixed numbers of four-simplices of
type (4,1), for $N_4^{(4,1)}=$ 40, 80
and 160k (left to right) at $\kappa_0\equ 2.2$ and $\Delta\equ 0.6$.
}}
\label{volumes}
\end{figure}

In order to obtain an efficient sampling of geometries of large
four-volume (the larger the four-volume, the better
the computer-generated geometries describe the continuum limit),
we have performed the simulations at (approximately) constant
four-volume. This means that instead of the standard gravitational
path integral (after Euclideanization) 
\beq
Z_E(\lambda,G) = \int \cD [g] \; e^{-S_E[g]},~~~
S_E[g]= -\frac{1}{G}\int d^4x \sqrt{|\det g|}\,(R-2\lambda)
\label{2.1}
\eeq
we simulate a discretized version of the path integral at
fixed four-volume $V_4$,
\beq
\tilde Z_E(V_4,G) =\! \int\!\cD [g] \, e^{-\tilde S_E[g]}\,
\del(\! \int\! d^4x \sqrt{\det g}-V_4),~~~
\tilde S_E[g]= -\frac{1}{G}\!\int\! d^4 x \sqrt{\det g}\,R,
\label{2.1b}
\eeq
which is related to \rf{2.1} by a Laplace transformation,
\beq\label{2.1a}
Z_E(\lambda,G) = \int_0^\infty dV_4\; e^{-\frac{1}{G} \lambda V_4}
\tilde Z_E(V_4,G).
\eeq
However, the local moves do not in general preserve the numbers
$N_4^{(4,1)}$ and $N_4^{(3,2)}$ of four-simplices, or indeed their
sum. We deal with this in a standard way which was
developed for dynamically triangulated models in
dimensions three and four \cite{av,aj}
to ensure that the system volume is peaked at a prescribed value,
with a well-defined range of fluctuations.
Adapting it to the current
situation of causal triangulations with nonvanishing asymmetry
$\Delta$, we
implement an approximate four-volume constraint by
adding a term
\beq
\delta S = \epsilon |N_4^{(4,1)}- \tilde N_4|,
\label{fixvolume}
\eeq
to the Euclidean action, with typical values of $\epsilon$ lying in
the range of 0.01 to 0.02, except during thermalization where we
set $\epsilon\equ 0.05$. The reason for fixing $N_4^{(4,1)}$
instead of $N_4\equ N_4^{(4,1)}\pl N_4^{(3,2)}$ in eq.\ \rf{fixvolume}
is mere technical convenience. 
We have checked in the phase space 
region relevant to four-dimensional quantum gravity 
(phase C, see below) that for $N_4^{(4,1)}$ fixed according
to \rf{fixvolume}, the number $N_4^{(3,2)}$ of four-simplices
of type (3,2) is likewise very sharply peaked, see Fig.\ \ref{volumes}.
\begin{table}
\begin{center}
\renewcommand{\arraystretch}{1.4}
\begin{tabular}{ |c||c|c|c|c|c|}
\hline
``four-volume" $\tilde N_4= N_4^{(4,1)}$  &   10 &   20 &  40 & 80 & 160 \\
\hline
actual four-volume $N_4= N_4^{(4,1)}\pl N_4^{(3,2)}$  & 22.25  &  45.5
& 91 & 181 & 362   \\
\hline
\end{tabular}
\end{center}
\caption{\label{ntab} Translation table between the two types of discrete
four-volume, $\tilde N_4$ and $N_4$, at which the numerical simulations
reported in this article were performed, in units of 1000 building blocks (phase C only).
}
\end{table}
The ``four-volumes" $\tilde N_4$ and the corresponding ``true"
discrete four-volumes $N_4$ used in the simulations are
listed in Table \ref{ntab}.
In order to stabilize the total volume after thermalization,
$\kappa_4$ has to be fine-tuned to its pseudo-critical value
(which depends weakly on the volume) with accuracy smaller
than $\epsilon$, in practice to about $0.2\epsilon$.
The measurements reported in this paper were taken
at $\tilde N_4\equ$ 10, 20, 40, 80 and 160k, and the runs
were performed on individual PCs or a PC farm for the smaller
systems and a cluster of work stations for the larger systems.

Before measurements can be performed, one needs  a well
thermalized configuration of a given volume. In order to
double-check the quality of the thermalization, we used two
different methods
to produce starting configurations for the measurement runs.
In the first method, we evolved from an
initial minimal four-dimensional triangulation of prescribed topology
and of a given time extension $t$, obtained by
repeated gluing of a particular triangulated space-time slice
of $\Delta \tau\equ 1$ and topology $[0,1]\times S^3$,
which consists of 30 four-simplices. The spatial in- and out-geometries
of the slice are minimal spheres $S^3$, made of five tetrahedra.
The two types of spatial boundary conditions used are (i) periodic
identification of the geometries at initial and final integer times,
and (ii) free boundary conditions, where all vertices contained
in the initial slice at time $\tau_0$ are
connected by timelike edges to a single vertex at time $\tau_0\mi 1$,
and similarly for the vertices contained in the final spatial slice.
From this initial configuration, the geometry evolves to its
target volume $\tilde N_4$ as specified through $\delta S$. During the
evolution the volume-volume correlator (see eq.\ \rf{scscs1} below)
changes from a very broad distribution to the stable narrow
shape seen in Fig.\ \ref{volvol}. The number of sweeps to reach the
thermalized configuration changes linearly with $\tilde N_4$ and
ranged from $10^5$ to $10^8$ sweeps for the largest volumes,
the latter of which took several weeks on a work station.

In the second method, we started instead from a
thermalized configuration of smaller volume, which we let
evolve towards the target volume. In this case the final
volume-volume distribution is reached from a narrower
distribution, namely, that of the smaller volume. During
thermalization, this width grows very slowly. The timing of
the entire process is similar to that of the first method.

\begin{figure}[t]
\centerline{\scalebox{0.6}{\rotatebox{0}{\includegraphics{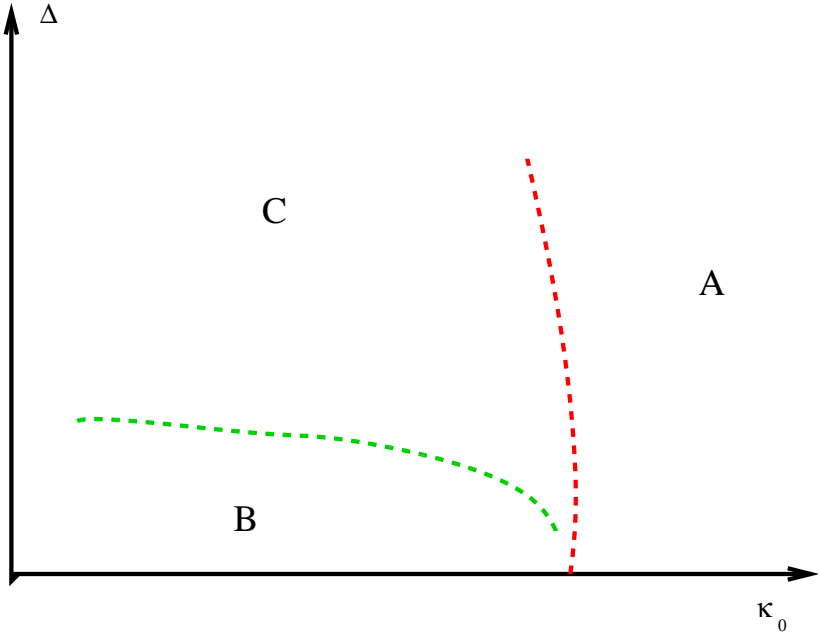}}}}
\caption[phased]{{\small A qualitative sketch of the phase diagram of four-dimensional causal
dynamical triangulations as function of the gravitational coupling $\kappa_0$
and the asymmetry parameter $\Delta$.}}
\label{fourdphasediagr}
\end{figure}

\section{Phase structure of the model}\label{phase}

\begin{figure}[t]
\centerline{\scalebox{0.4}{\rotatebox{0}{\includegraphics{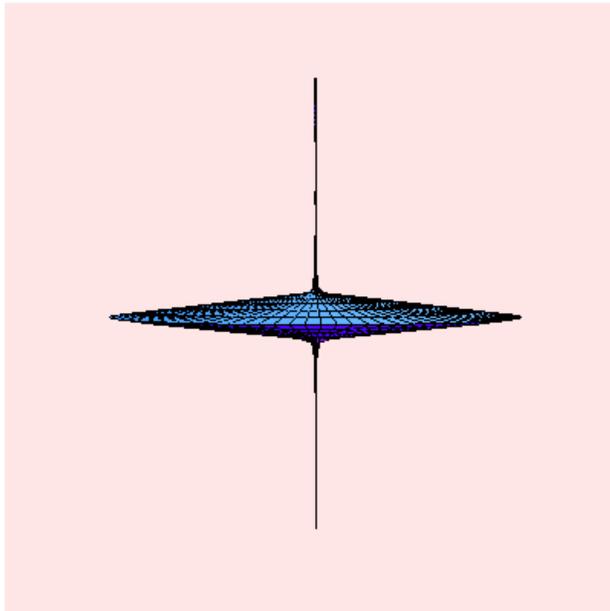}}}}
\caption[phased]{{\small
Monte Carlo snapshot of a typical universe in phase B ($\kappa_0\equ 1.6$),
of discrete volume $N_4\equ$ 22.25k
and total time extent $t\equ 20$. The entire universe has collapsed into a
slice of minimal time extension. }}
\label{uni1p6}
\end{figure}

\begin{figure}[ht]
\centerline{\scalebox{0.4}{\rotatebox{0}{\includegraphics{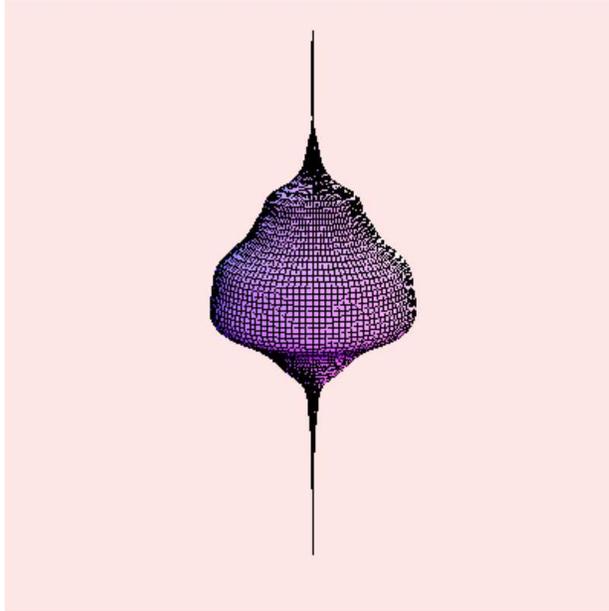}}}}
\caption[phased]{{\small
Monte Carlo snapshot of a typical universe in phase C ($\kappa_0\equ 2.2$)
of discrete volume $N_4\equ$ 91.1k
and total time extent $t\equ 40$.}}
\label{unipink}
\end{figure}

Like always in dynamically triangulated models, the bare cosmological
constant $\lambda^{(b)}$ (equivalently, $\kappa_4$) is tuned to its
(pseudo-)critical value in the
simulations, tantamount to approaching the infinite-volume limit.
Depending on the values of the two remaining parameters $\kappa_0$
and $\Delta$ in the discretized action \rf{actshort}, we have identified
three different phases, A, B and C, mutually separated by lines of
first-order transitions.
We have not yet measured their location in detail; Fig.\ \ref{fourdphasediagr}
is a qualitative sketch of the phase diagram. We describe each of the
three phases in turn:
\begin{figure}[ht]
\centerline{\scalebox{0.4}{\rotatebox{0}{\includegraphics{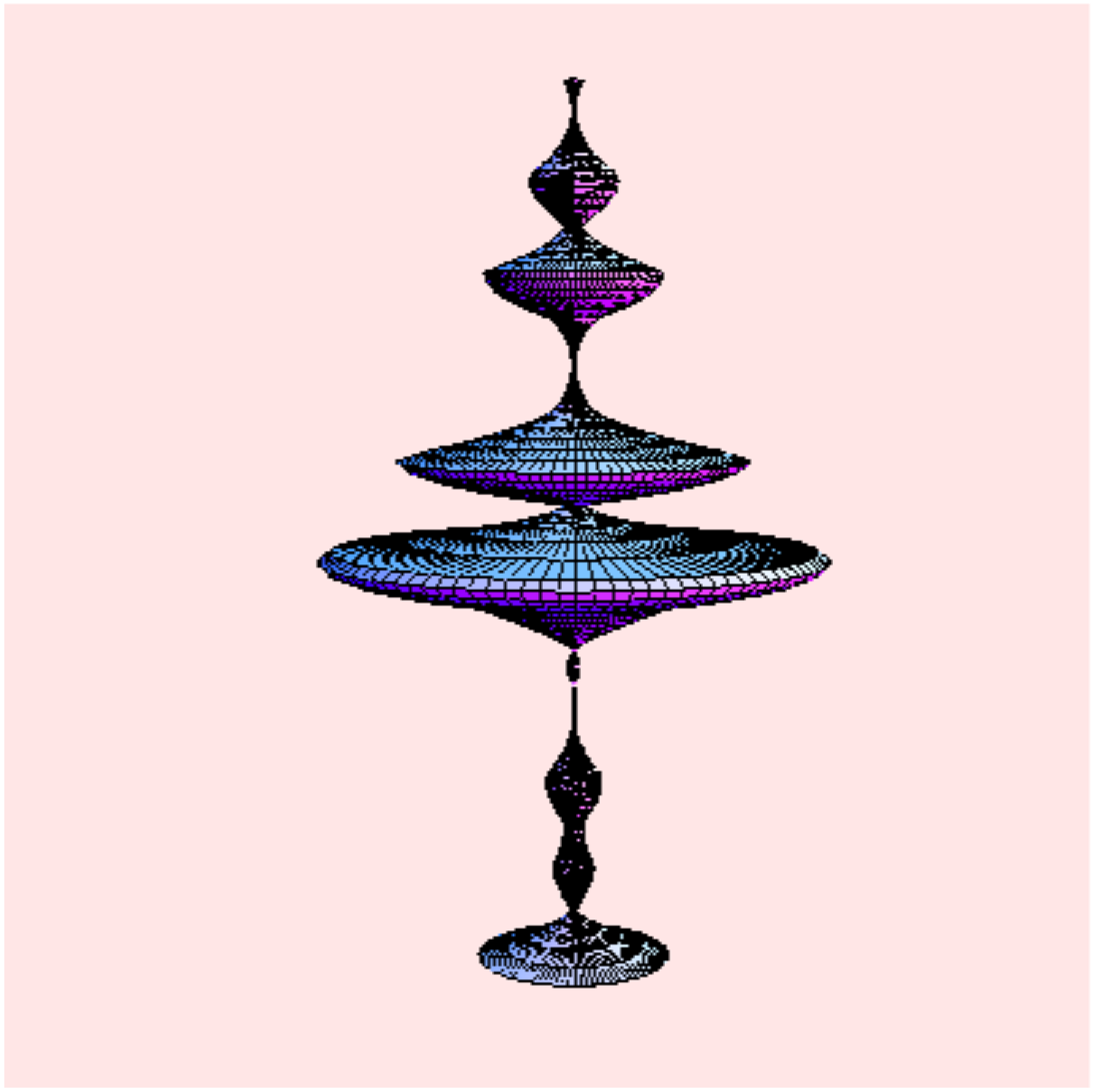}}}}
\caption[phased]{{\small
Monte Carlo snapshot of a typical universe in phase A ($\kappa_0\equ 5.0$, 
$\Delta\equ 0$),
of discrete volume $N_4\equ$ 45.5k
and total time extent (vertical direction) $t\equ 20$. In this and the following
two figures, the circumference at
integer proper time $\tau$ is chosen proportional to the spatial three-volume
$V_3(\tau)$. The surface represents an interpolation between
adjacent spatial volumes, without capturing the
actual 4d connectivity between neighbouring spatial slices.}}
\label{uni5p0}
\end{figure}
\begin{itemize}
\item[(A)] This phase prevails for sufficiently large $\kappa_0$
(recall $\kappa_0$ is proportional to the bare inverse Newton's constant
$k^{(b)}$).
When plotting the volume of the spatial slices $\tau \equ const$ as a function
of $\tau$, we observe an irregular sequence of maxima and minima,
where the minimal size is of the order of the cutoff, and the sizes of
the maxima vary, see Fig.\ \ref{uni5p0}. The time intervals during which
the spacetime has a macroscopic spatial extension are small and
of the order of $\Delta \tau\equ 3$.
\item[(B)] This phase occurs for sufficiently small $\kappa_0$ and
for small asymmetry $\Delta$, including $\Delta\equ 0$. In it,
spacetime undergoes a ``spontaneous dimensional reduction"
in the sense that all four-simplices are concentrated in a slice of
minimal time extension $\Delta \tau\equ 2$, and
the three-volume $N_3(\tau)$ remains close to its kinematic minimum
everywhere else (Fig.\ \ref{uni1p6}).
\end{itemize}
We believe that these two phases can be understood essentially from
the dynamics of the (Euclidean) geometries of the slices of constant
$\tau$ alone. The spatial slices in phase A are presumably realizations
of branched polymers which dominate the path integral of
three-dimensional Euclidean quantum gravity for large $\kappa_0$,
whereas the extended part of the geometry in phase B
corresponds to a Euclidean geometry in the so-called crumpled
phase, which is characterized by the presence of very few
vertices with extremely high coordination number \cite{av,abkv,simon,simon2}.
We will not elaborate here on our detailed measurements 
which quantify these statements,
because we view phases A and B as lattice artifacts.
However, there is yet another phase where the dynamics is not
reduced to just the spatial directions but is genuinely
four-dimensional.
\begin{itemize}
\item[(C)] This phase occurs for sufficiently small $\kappa_0$ and
nonvanishing asymmetry $\Delta$. In this phase, 
where $\Delta >0$ and $\tilde{\alpha}<1$ ($\tilde{\alpha}$ was defined 
below eq.\ \rf{act4dis}),
there seems to be a sufficiently strong coupling between successive
spatial slices to induce a change in the spatial
structure itself. We will discuss
this new geometrical structure in detail in Sec.\ \ref{thin}. 

Of course, the presence of the asymmetry
parameter $\Delta$ is only possible because of the Lorentzian character of
the model, and the resulting distinction between spatial and
time directions. Since we implemented this distinction in the
discretization by setting up a particular proper-time foliation 
and choosing arbitrary discrete length and time scales,
it should not be surprising that in order to make contact
with continuum physics, the ratio of these two
scales cannot be chosen completely arbitrarily. 
This merely implies a restriction on the way the discretization
is set up, and does not lead to any new parameters in
the continuum theory. We have checked that the physical results 
derived within phase C do not depend on the numerical
value of $\Delta$, within the allowed range. As
we will see later, any such choice leads to the
large-scale geometry of a homogeneous and
isotropic universe. 

The hallmark of this phase is the presence of
a stable extended four-geometry, as first reported in \cite{ajl-prl},
and illustrated in Fig.\ \ref{unipink}. Not only does the macroscopic
dimensionality of spacetime emerge with the correct classical value,
but, equally remarkably, the global shape of the universe has
been shown to be related to a simple minisuperspace action
similar to those used in standard quantum cosmology \cite{semi}.
Details of the computer measurements in this phase are the
subject of Secs.\ \ref{measure}-\ref{thick} below.
\end{itemize}

Looking at the phase diagram of Fig.\ \ref{fourdphasediagr}, it may at 
first be surprising that one does not obtain a continuum limit 
when the bare gravitational coupling constant (or Newton's constant)
vanishes, $G^{(b)}\sim 1/k^{(b)} \to 0$.
However, one should bear in mind that the
method of dynamical triangulation is a way of
defining the path integral, rather than a
regularization method which has nice smooth
continuum manifolds as its dominant configurations
as $G^{(b)} \to 0$. 

A helpful analogy is the example of the (Euclidean) path integral
for the relativistic particle, whose propagator $G(x,y)$ can
be represented as a sum over all random paths from $x$ to $y$, whose
action $S(P)\equ m^{(b)} L(P)$ is given by the length $L$ of the path $P$
in target space, multiplied by a dimensionless bare mass parameter $m^{(b)}$.
If this path integral is regularized on
a hypercubic lattice in $R^d$, the possible paths are by definition
along the lattice links. Now consider $R^2$ and a path from $(0,0)$ to
the lattice point $(x,y)\equ  (a \cdot n,a\cdot  n)$.
No matter how small the lattice spacing $a$ and how large the number $n$ of
lattice steps, the lattice paths between
the two points will never approach smooth continuum paths and
their lengths will always be larger than
$\sqrt{2}  \sqrt{x^2+y^2}$. To define a continuum limit
for the lattice random walk one looks for critical points of the
corresponding statistical system and studies the
long-distance properties of suitable observables, in this case, 
of the propagator.
The critical point is determined by the value of $m^{(b)}$, but
this bare coupling constant is usually not related to the physical
parameter (given here by the mass term in the continuum propagator) in a 
straightforward way (see \cite{book,rw}) for details). In the case of 
dynamical triangulations we follow a similar
strategy. We do not {\it a priori} associate a continuum limit with
certain values of the bare coupling constants. Rather, we look
for points or regions in coupling-constant space where interesting 
scaling limits may exist. In the case of four-dimensional gravity,
it is natural to look for regions where some
desirable features of macroscopic spacetime are reproduced,
and this is precisely the case for the phase-space region (C).

Let us also comment on how the continuum limit is taken in our
approach. According to the standard, Wilsonian understanding 
of the relation between
the theory of critical phenomena in statistical mechanics
and quantum field theory, in order to obtain a continuum quantum
field theory from an underlying regularized statistical system
one needs to fine-tune a few coupling constants (corresponding to
the relevant directions in coupling constant space) to reach
the infinite-dimensional critical surface where a suitably defined
correlation length is infinite. Directions along the critical surface
correspond to the infinitely many irrelevant (nonrenormalizable) directions
in coupling constant space. Weinberg conjectured that this picture,
which is usually believed to define ordinary renormalizable
quantum field theories, may generalize to quantum gravity,
in the sense that it may have an associated, nonperturbatively 
defined critical surface \cite{weinberg}. Several groups have since
tried to substantiate this idea using either a 
$(2\pl \epsilon)$-expansion \cite{kawai} or 
refined renormalization group techniques
\cite{reuteretc}. Various lattice approaches to quantum
gravity like quantum Regge calculus \cite{hamber}, Euclidean
dynamical triangulations \cite{aj,euclidean,aj1} and
the causal dynamical triangulation of quantum gravity 
\cite{ajl-prl,semi,ajl4d} used by us 
presently also endorse this picture. 

In the latter, Lorentzian approach, the bare cosmological constant 
must be fine-tuned to obtain an infinite spacetime volume. 
Formally, we are not performing an analogous fine-tuning 
of the bare gravitational coupling constant in phase C, which
is nonstandard from a Wilsonian point of view.
There {\it are} indeed lattice theories where no fine-tuning is 
needed, and where an infinite-volume limit is sufficient to identify
a continuum limit, for example, the Coulomb phase of $U(1)$-lattice gauge
theory. However, the situation in gravity could be more involved,
and subtleties may be buried in the identification of more complicated
observables than the global ones we have considered so far. 
For a related discussion in three-dimensional causal dynamically
triangulated quantum gravity, we refer to \cite{ajlabab,ajlv}. 
What is important from our present point of view is the fact that 
the universes emerging from phase C satisfy the first
nontrivial tests of being viable candidates for quantum universes,
as we will now turn to describe. 
Encouraged by these results, we are in the process of developing 
new and more refined
methods to probe the geometry of the model further, and which eventually 
should allow us to test aspects related to its local ``transverse" degrees 
of freedom, the
gravitons. We invite and challenge our readers to find such tests in a
truly background-independent formalism of quantum gravity.

\section{Measuring geometry}\label{measure}

Having generated a ground state of quantum geometry
nonperturbatively, we would like to understand its geometric
properties, in the sense of expectation values, which can be done
in various ways. We will currently concentrate on the purely
geometric observables, leaving the coupling to test particles
and matter fields to a later investigation.
We will proceed by first determining a
number of ``rough" properties of the quantum geometry
such as its dimensionality and its global scaling properties.
These properties of quantum spacetime
play an important role, because for any viable candidate
theory of quantum gravity, they must be shown to
reproduce the correct classical limit at sufficiently large scales,
namely, geometry as
described by Einstein's general theory of relativity.

As we have emphasized in \cite{ajl-prl}, in nonperturbative
models of quantum gravity, where by definition geometry
fluctuates very strongly at short scales, there is absolutely
no guarantee that any sensible classical limit is obtained even
for very basic quantities like the dimension of spacetime.
Indeed, previous formulations which can probe the
dimensionality of geometry numerically
through the scaling behaviour of suitable observables, for example,
dynamically triangulated models of Euclidean quantum gravity,
have not come up with evidence for the value 4.
Instead such models seem to run generically into highly
degenerate geometries of one type or other \cite{aj1,high-vertex,bielefeld2,simon2},
which at large scales have nothing to do with four-dimensional spacetime
as we know it. From this
it seems likely that the condition of
obtaining the correct classical limit imposes strong constraints
on any attempt to quantize gravity
{\it nonperturbatively and background-independently}.
Of course, proposals of quantizing gravity are often not
sufficiently developed to decide whether they can reproduce
classical geometry from first principles, without putting in a
preferred background structure along the way.

In the following, we will present a detailed discussion of a
number of geometric properties of the extended ground state
of phase C above. Several main results were already announced
in our previous publications \cite{ajl-prl,semi,newpaper}. The picture
emerging so far is that of a geometry whose large-scale
properties are classical, but whose detailed microscopic
structure is highly complicated and nonclassical. We will
proceed by first presenting the evidence for the macroscopic
four-dimensionality of spacetime, and then describing the
insights into the geometric structure obtained from measuring
quantities associated with slices of constant time.

\subsection{The evidence for spacetime dimension four: \\ Global scaling}

As illustrated by Fig.\ \ref{unipink}, the total four-volume is
distributed over the $t$ time steps (we are working with a
periodic identification in the time direction) in a characteristic
way. At a given discrete four-volume, for sufficiently large $t$
the configuration separates out into a thin ``stalk" whose
four-volume at each time step fluctuates little and
is close to the kinematically
allowed minimum for the topology $[0,1]\times S^3$, and
an extended, blob-shaped structure, the ``universe".
The average volume
$s$ per time step $\Delta \tau\equ 1$ of the stalk is practically
independent of the total volume of the system, with
$s\approx 25.5$ for the range of coupling constants we
used. As explained in Sec.\ \ref{num}, we measure the
spacetime volume by $\tilde N_4$, that is, 
in units of four-simplices of type (4,1).
Since every spatial tetrahedron
is contained in exactly two (4,1)-simplices, the average
three-volume $N_3(\tau)$ at integer time $\tau$
in the stalk is therefore given by $s/2$.

\begin{figure}[ht]
\vspace{-3cm}
\psfrag{tau}{\bf{\Large $x$}}
\psfrag{VV}{\Large\bf $c_{\tilde N_4}(x)$}
\centerline{\scalebox{0.6}{\rotatebox{0}{\includegraphics{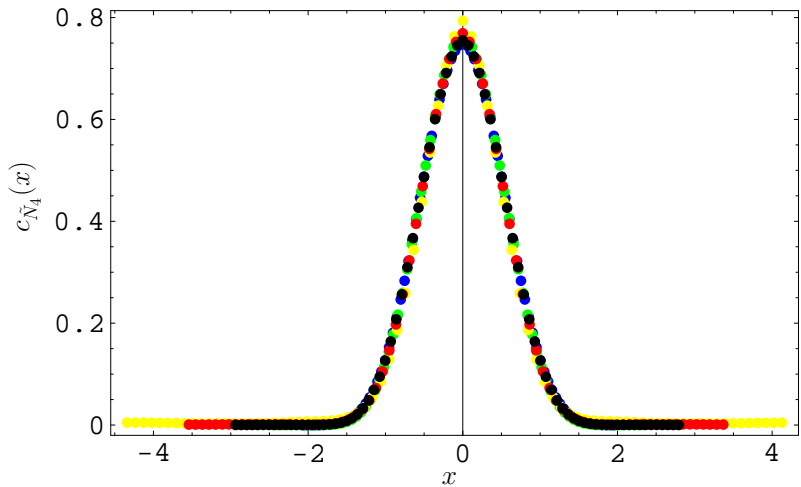}}}}
\vspace{-4.5cm}
\caption[phased]{{\small The scaling of the volume-volume distribution,
as function of the rescaled time variable $x\equ \delta/(\tilde N_4^{\it eff})^{1/4}$.
Data points come from system sizes $\tilde N_4\equ$ 10, 20, 40, 80 and 160k at
$\kappa_0\equ 2.2$, $\Delta\equ 0.6$ and $t\equ 80$.}}
\label{scscs1}
\end{figure}

\begin{figure}[ht]
\vspace{-3cm}
\psfrag{d}{{\Large $D_H$}}
\psfrag{err}{\Large error}
\centerline{\scalebox{0.6}{\rotatebox{0}{\includegraphics{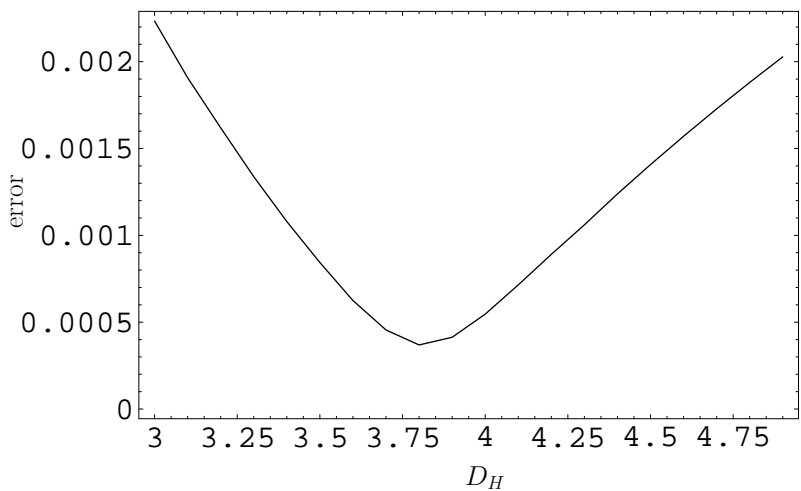}}}}
\vspace{-4.5cm}
\caption[phased]{{\small Estimating the best scaling dimension $D_H$
from optimal overlaps of rescaled volume-volume distributions.
The plot variable is the average error of the overlap as function of $D_H$.}}
\label{err}
\end{figure}

A suitable observable for studying the large-scale structure
of the universe is the volume-volume correlator\footnote{Note
that in the analogous relation (6) in \cite{ajl-prl} we mistakenly
omitted the ensemble averages on the right-hand side.}
\beq
C_{\tilde N_4}(\delta) \equiv \langle N_3(0) N_3(\delta)\rangle=
\sum_{\tau\equ 1}^t \frac{4\langle (N_3(\tau)-s/2)(N_3(\tau+\delta)-s/2)
\rangle}{(\tilde N_4-t s)^2},
\label{volvol}
\eeq
which satisfies the normalization condition
\beq
\sum_{\delta\equ 0}^{t-1}C_{\tilde N_4}(\delta)  = 1,
\label{norma}
\eeq
where a periodic identification of the time variable has
been assumed. 
Because of the subtraction of the cutoff
volume $s/2$, the volume-volume correlator is only sensitive
to the extended part of the geometry, and measures the
distribution of the ``effective" four-volume
\beq
\tilde N_4^{\it eff}=\tilde N_4-t s
\label{veff}
\eeq
contained in it.
We have measured the distribution \rf{volvol} for system sizes
$\tilde N_4\equ$ 10, 20, 40, 80 and 160k,
at $\kappa_0\equ 2.2$, $\Delta\equ 0.6$ and $t\equ 80$.
We expect the distribution to scale as a function of $\tilde N_4$,
with an underlying universal distribution
\beq
c_{\tilde N_4}(x):=(\tilde N_4^{\it eff})^{1/D_H} C_{\tilde N_4
}( (\tilde N_4^{\it eff})^{1/D_H}x )
\label{volvolre}
\eeq
depending on the rescaled time variable\footnote{We are employing
standard finite-size scaling methods familiar from the theory
of critical phenomena (see \cite{barkema} for a textbook exposition),
adjusted to our problem.}
\beq
x=\frac{\delta}{(\tilde N_4^{\it eff})^{1/D_H}}.
\label{xdef}
\eeq
The value of the scaling dimension $D_H$ is then determined
by the method of ``best overlap". In it, we compare each measured
distribution with that of the 40k system. In order to provide a
comparison point for each value of $x$ and to quantify the overlap,
we first performed a spline
interpolation of the discrete distribution at 40k. Fig.\ \ref{scscs1} shows a
combined plot of the rescaled distributions,
using $D_H\equ 4$ as a scaling dimension, and with a
plot range symmetrized around $x\equ 0$.
The overlap is nearly perfect. As can be read off Fig.\ \ref{err}, which shows
the average error of the overlap as a function of $D_H$,
the best overlap corresponds to $D_H\approx 3.8\pm 0.25$.
We take this outcome as strong evidence that our spacetimes
are genuinely four-dimensional at large scales.\footnote{Of course,
we will never be able to show by computer simulations
alone that the effective dimension is exactly equal to four. What we
are assuming here is that this dimension is indeed an integer,
because there are no classical theories describing the dynamics
of geometries of dimensionality 3.9, say. Also, it is known from the
measurement of Hausdorff dimensions in two-dimensional Euclidean
quantum gravity (where it can be proven that the Hausdorff dimension is
four (!) \cite{kawai-watabiki,aw}) that the computer simulations
tend to slightly underestimate the actual values \cite{d_h-syracuse,ajw}.}

\subsection{The evidence for spacetime dimension four:\\ Spectral dimension}
\label{fourspectral}

Another way of obtaining an effective dimension of the
nonperturbative ground state, its so-called {\it spectral dimension} $D_S$,
is from studying a diffusion
process on the underlying geometric ensemble. On a $d$-dimensional 
manifold with a fixed,
smooth Riemannian metric $g_{ab}(\xi)$, the diffusion equation has the form
\beq\label{ja2}
\frac{\prt}{\prt \sg} \, K_g(\xi,\xi_0;\sg) = \Del_g K_g (\xi,\xi_0;\sg),
\eeq
where $\sg$ is a fictitious diffusion time, $\Del_g$ the Laplace
operator of the metric $g_{ab}(\xi)$ and $K_g(\xi,\xi_0;\sg)$
the probability density of diffusion from $\xi$ to $\xi_0$ in
diffusion time $\sg$. We will consider diffusion processes which 
initially are peaked at some point $\xi_0$,
\beq\label{ja3}
K_g(\xi,\xi_0;\sg\equ 0) = \frac{1}{\sqrt{\det g(\xi)}}\, \del^d(\xi-\xi_0).
\eeq
For the special case of a flat Euclidean metric, we have
\beq\label{flat1}
K_g(\xi,\xi_0;\sg) = \frac{\e^{-d_g^2(\xi,\xi_0)/4\sg}}{(4\pi \sg)^{d/2}},
\qquad g_{ab}(\xi)\equ \delta_{ab}.
\eeq
For general curved spaces $K_g$ has the well-known
asymptotic expansion
\beq\label{ja4}
K_g(\xi,\xi_0;\sg) \sim \frac{\e^{-d_g^2(\xi,\xi_0)/4\sg}}{\sg^{d/2}}
\sum_{r\equ 0}^\infty a_r(\xi,\xi_0)\, \sg^r
\eeq
for small $\sg$, 
where $d_g(\xi,\xi_0)$ denotes the geodesic distance
between $\xi$ and $\xi_0$.
Note the appearance of the power $\sg^{-d/2}$ in this relation, reflecting
the dimension $d$ of the manifold, just like in the formula \rf{flat1} for
flat space. This happens because small $\sg$'s correspond to short
distances and for any given smooth metric short distances imply 
approximate flatness.

A quantity that is easier to measure in numerical simulations is the average 
{\it return probability} $P_{g}(\sg)$, which possesses an analogous expansion
for small $\sg$,
\beq\label{ja5}
P_{g}(\sg) \equiv \frac{1}{V} \int  d^d\xi \sqrt{\det g(\xi)} \; K_g(\xi,\xi;\sg) \sim
\frac{1}{\sg^{d/2}} \sum_{r\equ 0}^\infty A_r \sg^r,
\eeq
where $V$ is the spacetime volume $V\equ\int d^d\xi \sqrt{\det g(\xi)}$ and
the expansion coefficients $A_r$ are given by
\beq\label{adef}
A_r = \frac{1}{V}\int  d^d\xi \sqrt{\det g(\xi)} \;a_r(\xi,\xi).
\eeq
For an infinite flat space, we have $P_g(\sg)\equ  1/(4\pi\sg )^{d/2}$ and thus can
extract the dimension $d$ by taking the logarithmic derivative,
\beq\label{ja5a}
-2\ \frac{d \log P_g(\sg)}{d\log \sg} = d,
\eeq
independent of $\sg$. For nonflat spaces and/or finite volume $V$,
one can still use
eq.\ \rf{ja5a} to extract the dimension, but there will be
corrections for sufficiently large $\sg$.
For finite volume in particular, $P_g(\sg)$ goes to $1/V$ for
$\sg \gg V^{2/d}$ since the zero mode of the Laplacian $-\Del_g$
will dominate the diffusion in this region. 
For a given diffusion time $\sg$ the behaviour of $P_g(\sg)$ is determined by
eigenvalues $\lambda_n$ of $-\Del_g$ with $\lambda_n \le 1/\sg$,
and the contribution from higher eigenvalues is exponentially suppressed.
Like in the flat case, where diffusion over a time $\sg$ probes the geometry 
at a linear scale $\sqrt{\sg}$, large $\sg$ corresponds to large distances away
from the origin $\xi_0$ of the diffusion process, and small $\sg$ to short distances.

The construction above can be illustrated by the simplest example of
diffusion in one dimension. 
The solution to the diffusion equation on the real axis is
\beq\label{jb1}
K(\xi,\sg) = \frac{\e^{-\xi^2/4\sg}}{\sqrt{4\pi \sg}},~~~~
K(k,\sg) = \e^{-k^2 \sg},
\eeq
where $K(k,\sg)$ denotes the Fourier transform of $K(\xi,\sg)$.
The eigenvalues of the Laplace operator are of course just given by $k^2$.
In order to illustrate the finite-volume effect, let us compactify 
the real line to a circle of length $L$. The return probability is now given by
\beq\label{jb2}
P_L(\sg)= \frac{1}{L}\sum_{n=-\infty}^\infty \e^{-k_n^2 \sg} =
\frac{1}{\sqrt{4\pi \sg}}
\sum_{m=-\infty}^{\infty} \e^{-L^2m^2/4\sg},~~~~k_n=\frac{2\pi n}{L},
\eeq
where in the second step we have performed a Poisson resummation to
highlight the $\sg^{-1/2}$-dependence for small $\sg$. In line with the
discussion above, the associated spectral dimension $D_S(\sg)$ 
is constant (and equal to one) up to
$\sg$-values of the order $L^2/4\pi^2$, and then goes to zero monotonically.

In applying this set-up to four-dimensional quantum gravity in a path integral formulation,
we are interested in measuring the expectation value of the return
probability $P_g(\sg)$. Since $P_g(\sg)$ defined in \rf{ja5}
is invariant under reparametrizations, it
makes sense to take its quantum average over all geometries of
a given spacetime volume $V_4$,
\beq\label{ja7}
P_{V_4}(\sg) =
\frac{1}{\tilde Z_E(V_4)} \int\!\! \cD [g_{ab}] \; e^{-\tilde{S}_E(g_{ab})}
\del(\int d^4x \sqrt{\det g}-V_4) \, P_g(\sg),
\eeq
where the partition function $\tilde Z_E$ and the corresponding gravitational
action $\tilde S_E$ were defined in \rf{2.1b}.
Since the small-$\sg$ behaviour of $P_g(\sg)$ is the same for each smooth 
geometry, it might seem obvious that the same is true for their integral $P_{V_4}(\sg)$,
but this need not be so. Firstly, the scale $\sg$ in \rf{ja7} is held fixed,
independent of the geometry $g_{ab}$, while the expansion \rf{ja5} contains
reference to higher powers of the curvature of $g_{ab}$.
Secondly, one should keep in mind that a typical geometry which contributes
to the path integral -- although continuous -- is unlikely to be smooth.
This does not invalidate our treatment, since diffusion processes can be
meaningfully defined on much more general objects than smooth
manifolds. For example, the return probability for diffusion on
fractal structures is well studied in statistical physics and takes
the form
\beq\label{ja6}
P_N(\sg)= \sg^{-D_S/2} \; F\Big(\frac{\sg}{N^{2/D_S}}\Big),
\eeq
where $N$ is the ``volume'' associated with the fractal
structure and $D_S$ the so-called {\it spectral dimension},
which is not necessarily an integer.
An example of fractal structures are branched polymers (which
will play a role in Sec.\ \ref{thin} below when we analyze the geometry of 
spatial slices), which generically have $D_S\equ 4/3$
\cite{thordur-john,anrw}. 
Extensive numerical simulations \cite{2dspectral,aa}
have shown that in 2d quantum gravity
the only effect of integrating over geometries is to replace
the asymptotic expansion \rf{ja5}, which contains
reference to powers of the curvature related to
a specific metric, by the simpler form \rf{ja6}.

Our next task is to define diffusion on the class of metric spaces
under consideration, the piecewise linear structures
defined by the causal triangulations $T$.
We start from an initial probability distribution 
\beq
K_T(i,i_0;\sg\equ 0) = \delta_{i,i_0},
\eeq
which vanishes everywhere except at a randomly chosen (4,1)-simplex $i_0$,
and define the diffusion process by the evolution rule
\beq
K_T(j,i_0;\sg+1) = \frac{1}{5}\sum_{k\to j} K_T(k,i_0;\sg).
\label{evo4d}
\eeq
These equations are the simplicial analogues of \rf{ja3} and \rf{ja2},
with the triangulation (together with its Euclideanized edge-length assignments)
playing the role of $g_{ab}$, and
$k\to j$ denoting the five nearest neighbours of the four-simplex $j$.
In this process, the total probability
\beq
\sum_j K_T(j,i_0;\sg) =1
\eeq
is conserved. The return probability to a simplex $i_0$ is then
defined as $P_T(i_0;\sg)\equ  K_T(i_0,i_0;\sg)$ and the
quantum average as
\beq\label{ja7a}
P_{N_4}(\sg)= \frac{1}{\tilde Z_E(N_4)} \sum_{T_{N_4}} 
e^{-\tilde{S}_E(T_{N_4})}\;
\frac{1}{N_4}\sum_{i_0 \in T_{N_4}} K_{T_{N_4}}(i_0,i_0;\sg),
\eeq
where $T_{N_4}$ denotes a triangulation with $N_4$ four-simplices,
and $\tilde{S}_E(T_{N_4})$ and $\tilde Z_E(N_4)$ are the obvious
simplicial analogues of the continuum quantities
\rf{2.1b} at fixed four-volume.
Assuming that the return probability behaves according to \rf{ja6},
with $N\equ N_4$, we can extract the value of the fractal dimension $D_S$ 
by measuring the logarithmic derivative as in \rf{ja5a} above, as long
as the diffusion time is not much larger than $N_4^{2/D_S}$,
\beq\label{ja1}
D_S(\sg) = -2 \;\frac{d\log P_{N_4}(\sg)}{d\log \sg}+\mbox{finite-size corrections}.
\eeq

From the experience with numerical simulations of 2d Euclidean quantum
gravity in terms of dynamical triangulations \cite{ajw,2dspectral,aa}, we
expect some irregularities in the behaviour of the return probability for
the smallest $\sg$, i.e. close to the cut-off scale. Typically, the
behaviour of $P_N(\sg)$ for odd and even diffusion steps $\sg$ will
be quite different for small $\sg$ and merge
only for $\sg \approx 20-30$. After the merger, the curve enters 
a long and stable regime where the right-hand side of \rf{ja1} is 
independent of $\sg$, before finite-size effects start to dominate
which force it to go to zero.
(Also the 3d Euclidean hypersurfaces of
constant time in our current set-up show a very similar behaviour,
as can be seen in Fig.\ \ref{ds2.2b4}.)

The origin of the odd-even asymmetry can again be illustrated by the
simple case of diffusion on a one-dimensional circle, whose
solution is the discretized version of the solution \rf{jb2}. In this
case the asymmetry of the return probability between odd and even time steps
is extreme: if we use the two-dimensional version
of the simple evolution equation \rf{evo4d}, we obtain 
\beq\label{jb3}
P_L(\sg)= \left\{
\begin{array}{cl}
0 & ~~\mbox{for $\sg$ {odd,}} \\
~&~\\
\displaystyle{\frac{1}{2^\sg}}\;
\pmatrix{\sg \cr {\sg}/{2}} &~~\mbox{ for $\sg$ even,}
\end{array}
\right.
\eeq
as long as $\sg < L/2$, where $L$ is the discrete volume of the circle
(i.e. the number of its edges).
It is of course possible to eliminate this asymmetry by using an
``improved" discretized diffusion equation, but in the case of
higher-dimensional random geometries like the ones used in 4d 
causal dynamical triangulations
this is not really necessary. 
The random connectivity will eliminate the asymmetry after a number
of steps as exemplified by Fig.\ \ref{ds2.2b4}.
\begin{figure}[t]
\psfrag{x}{{\bf{\LARGE $\sg$}}}
\psfrag{y}{{ \bf{\LARGE $D_S$}}}
\centerline{\scalebox{0.7}{\rotatebox{0}{\includegraphics{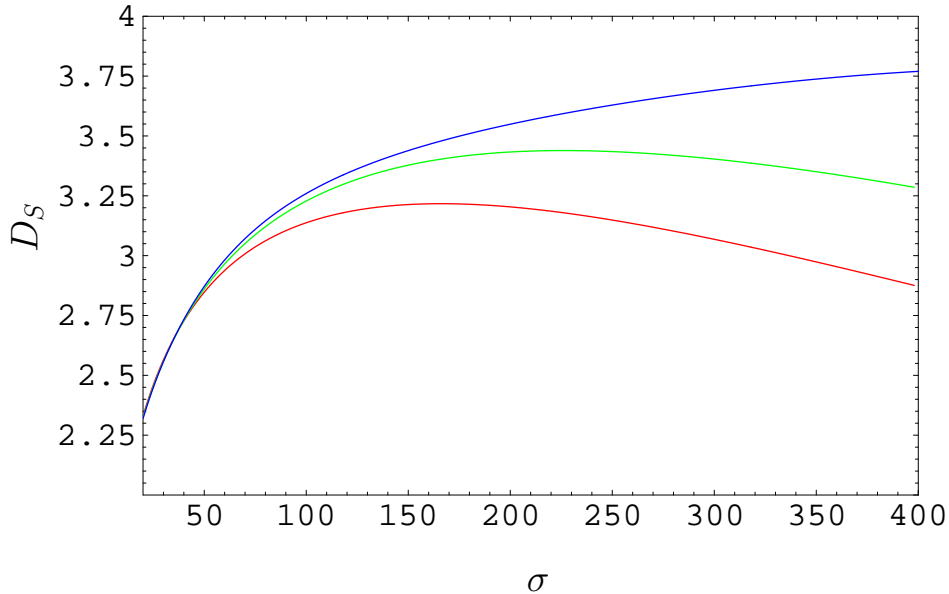}}}}
\vspace{-4.8cm}
\caption[phased]{{\small The spectral dimension $D_S(\sg)$ of causal
dynamical triangulations as a function of the diffusion time $\sg$, which
is a direct measure of the distance scale probed. The measurements  
were taken at volumes $\tilde N_4\equ 20$k (bottom curve), 40k and
80k (top curve), and for $\kappa_0\equ 2.2$, $\Delta\equ 0.6$ and 
$t\equ 80$.}}
\label{new4d2a}
\end{figure}

The results of measuring the spacetime spectral dimension $D_S$
were first reported in \cite{newpaper}. We work with system sizes of
up to $\tilde N_4\equ 80$k with $\kappa_0\equ 2.2$, $\Delta\equ 0.6$ and 
$t\equ 80$. Since we are interested in the bulk properties of quantum
spacetime and since the volume is not distributed evenly in the
time direction (cf. Fig.\ \ref{unipink}), 
we always start the diffusion process from a four-simplex
adjacent to the slice of maximal three-volume.
When this is done the
variations in the curve $D_S(\sg)$ for different
generic choices of $T_N$ and starting simplices $i_0$ are small, as long
as $i_0$ is located not too close to the stalk.
The data curves presented in Fig.\ \ref{new4d2a} were
obtained by averaging over 400 different diffusion processes
performed on independent configurations. 
Because following the diffusion
out to a time $\sg\equ 400$ is rather time-consuming and the
variation in $P_T(i_0;\sg)$ as a function of $T$ and $i_0$ (with
the initial conditions specified above) is small, we have not
performed a more extensive average over configurations.
We have omitted error bars from Fig.\ \ref{new4d2a} to illustrate
how the curves converge to a shape that represents $D_S(\sg)$ in 
the infinite-volume
limit, and which is given by the envelope of the data curves for finite volume.
For the two lower volumes, $\tilde N_4\equ 20$k and $\tilde N_4\equ 40$k,
there still are clear finite-volume effects for large diffusion times $\sg$.

\begin{figure}[t]
\psfrag{X}{{$\sg$}}
\psfrag{Y}{{ $D_S$}}
\centerline{\scalebox{1.2}{\rotatebox{0}{\includegraphics{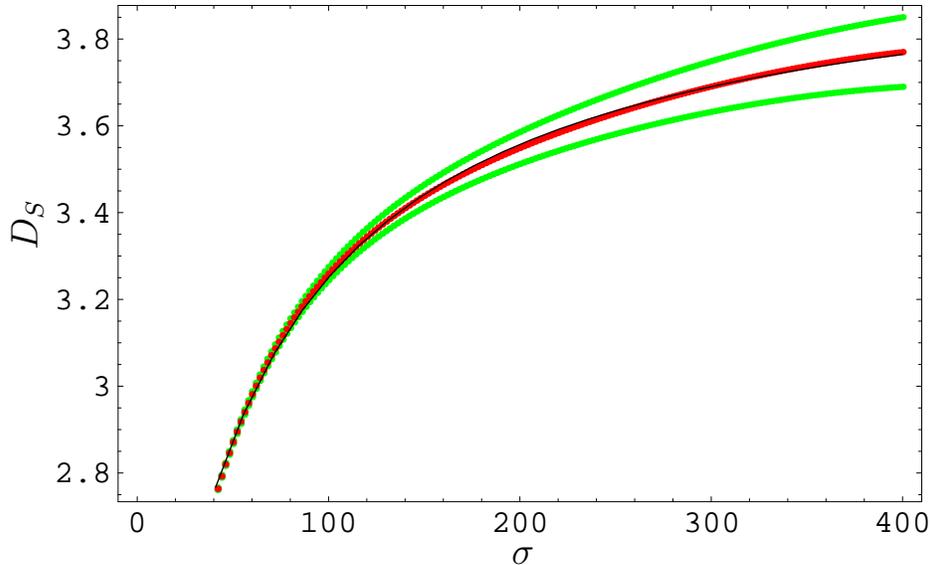}}}~~~~
~~~}
\caption[phased]{{\small The spectral dimension $D_S$ of the universe as
function of the diffusion time $\sg$, measured for
$\kappa_0\equ 2.2$, $\Delta\equ 0.6$ and $t\equ 80$, and a spacetime volume
$N_4\equ 181$k. The averaged measurements lie along the central curve, 
together with a superimposed best fit $D_S(\sg) = 4.02\mi 119/(54\pl\sg)$. The two outer
curves represent error bars.}}
\label{d4s2.2b4}
\end{figure}

By contrast, the top curve -- corresponding to the maximal volume 
$\tilde N_4\equ 80$k -- continues to rise for increasing $\sg$, which makes
it plausible that we can ignore any finite-size effects and that it is a good 
representative of the infinite-volume limit in the 
$\sg$-range considered.\footnote{In both Figs.\ \ref{new4d2a} and
\ref{d4s2.2b4} we have only plotted the region where the curves
for odd and even $\sg$ coincide, in order to exclude short-distance
lattice artifacts. The two curves merge at about $\sg\equ 40$.
Since the diffusion distance grows as $\sqrt{\sg}$, a
return diffusion time of 40 corresponds to just a few steps
away from the initial four-simplex.}
We will therefore concentrate on analysing the
shape of this curve, which is presented separately in Fig.\ \ref{d4s2.2b4},
now with error bars included. (More precisely, the two outer curves 
represent the envelopes to the tops and bottoms of the error bars.) The
error grows linearly with $\sg$, due to the occurrence of the log$\, \sg$ in \rf{ja1}.

The remarkable feature of the curve $D_S(\sg)$ is its slow approach to the
asymptotic value of $D_S(\sg)$ for large $\sg$. 
As emphasized in \cite{newpaper}, this type of behaviour has never been
observed previously in systems of random geometry, and again
underlines that causal dynamical triangulations in four dimensions
behave qualitatively differently, and that the quantum geometry produced
is in general richer. The new phenomenon we observe here is a
{\it scale dependence of the spectral dimension}, which has emerged
dynamically. This is to be contrasted with fractal structures which
show a self-similar behaviour at all scales, and which lead to 
$D_S(\sg)$-curves like that shown in Fig.\ \ref{ds2.2b4}.

As explained in \cite{newpaper}, the best three-parameter fit which
asymptotically approaches a constant is of the form
\beq\label{ja9}
D_S(\sg) =  a -\frac{b}{\sg+c} = 4.02-\frac{119}{54+\sg}.
\eeq
The constants $a$, $b$ and $c$ have been determined by using the 
full data range $\sg \in [40,400]$
and the curve shape agrees well with the measurements, as can be seen from 
Fig.\ \ref{d4s2.2b4}.
Integrating \rf{ja9} we obtain
\beq\label{ja8a}
P(\sg) \sim \frac{1}{\sg^{a/2} (1+c/\sg)^{b/2c}},
\eeq
from which we deduce the limiting cases
\beq\label{ja8b}
P(\sg) \sim \left\{
\begin{array}{cl}
\displaystyle{{\sg^{-a/2}}} &~~\mbox{for large $\sg$,}\\
~&~\\
\displaystyle{{\sg^{-(a-b/c)/2}}} &~~ \mbox{for small $\sg$}.
\end{array}
\right.
\eeq

We conclude that the quantum geometry generated by
causal dynamical triangulations has a scale-dependent 
spectral dimension which increases continuously from
$a\mi b/c$ to $a$ with increasing distance.
Substituting the values for $a$, $b$ and $c$ obtained from the
fit \rf{ja9}, and taking into account their variation as we vary the 
$\sg$-range $[\sg_{\rm min}, \sg_{\rm max}]$ and use different
weightings for the errors, we obtained
the asymptotic values \cite{newpaper}
\beq\label{ja10}
D_S(\sg\equ \infty) = 4.02 \pm 0.1
\eeq
for the ``long-distance spectral dimension" and
\beq\label{ja11}
D_S(\sg\equ 0)= 1.80 \pm 0.25
\eeq
for the ``short-distance spectral dimension". A dynamically generated
scale-dependent dimension with this behaviour is truly exciting news,
because it signals the existence of an effective ultraviolet cut-off for
theories of gravity with and without matter coupling,
brought about by the (highly non-perturbative) behaviour of the 
quantum-geometric degrees of freedom on the very smallest  
scale.

\subsection{The effective action}

In order to get a more detailed understanding of the geometric properties
of our universe, we next attempt
to reproduce the overall shape of the universe of
Fig.\ \ref{unipink} (in other words, the behaviour of its three-volume $V_3$
as a function of time) from an effective action for the single variable
$V_3(\tau)$, valid at sufficiently large scales $V_3\gg 1$. This will enable
us to compare our results with those from so-called minisuperspace
models. The simplest such models also take the form
of dynamical systems of a single variable, namely, the scale factor $a(\tau)$,
describing the linear spatial extension of the universe and related to the
three-volume by
\beq
V_3(\tau) \propto a^3(\tau).
\label{scalef}
\eeq
Despite the resemblance, the status of our derivation is very different
from that of a minisuperspace model. In the latter, one works from the
outset in a drastically truncated framework, where the classical
dynamics of the
entire universe is described by the variable $a(\tau)$ instead of the
full metric data $g_{ij}(\tau,x)$. For closed universes, and in (Euclidean)
proper-time coordinates, this corresponds to considering line elements of
the form
\beq
ds^2 = d\tau^2 + a^2(\tau) d\Omega_3^2,
\label{minimetric}
\eeq
where $d\Omega_3^2$ denotes the metric on the three-sphere.
The corresponding Einstein-Hilbert action\footnote{We work with Euclidean
signature in order to facilitate comparison with the simulation data,
which also describe Euclidean, Wick-rotated geometries.} is
\beq
S^{\rm mini}= \frac{1}{G} \int d\tau \left( -a(\tau)
\left(\frac{d a(\tau)}{d\tau}\right)^2
-a(\tau) + \lambda a^3 (\tau)\right),
\label{miniaction}
\eeq
including a cosmological constant $\lambda$.
These minisuperspace models are often taken as starting point
for a path integral quantization, leading to quantum cosmologies like those
used by Hartle and Hawking in their semiclassical
evaluation of the wave function of the universe \cite{hh}
(see also \cite{eqg,vilenkin,linde,rubakov,hl,haha,vilcos,vilrev} for related work).
The hope is that such a truncated version of quantum gravity still
describes some features of the full theory correctly.

By contrast, the central feature of dynamical triangulations is a
bona fide path integration
over {\it all} geometric degrees of freedom (which in particular will have a
nontrivial spatial dependence), without appealing to any symmetry
reduction, either classical or quantum-mechanical.
In the final result, one may then decide to keep track only of the
three-volume (equivalently, the scale factor) as a dynamical variable and
try to extract its effective large-scale behaviour.
It is an interesting question whether the two approaches (that is, truncating
the degrees of freedom before or after the quantization) are equivalent.
In view of the severe difficulties in making
quantum-cosmological path integrals well-defined and convergent \cite{haha}, one
might hope that they are not. This is indeed the conclusion from the findings
in causal dynamical triangulations we are about to describe. As we will
see, the collective effect of the full quantum-gravitational degrees
of freedom changes the dynamical description in terms of the single
scale factor in a drastic way.

\begin{figure}[ht]
\vspace{-3cm}
\psfrag{d}{\bf{\Large $D_2$}}
\psfrag{err}{\Large\bf error}
\centerline{\scalebox{0.6}{\rotatebox{0}{\includegraphics{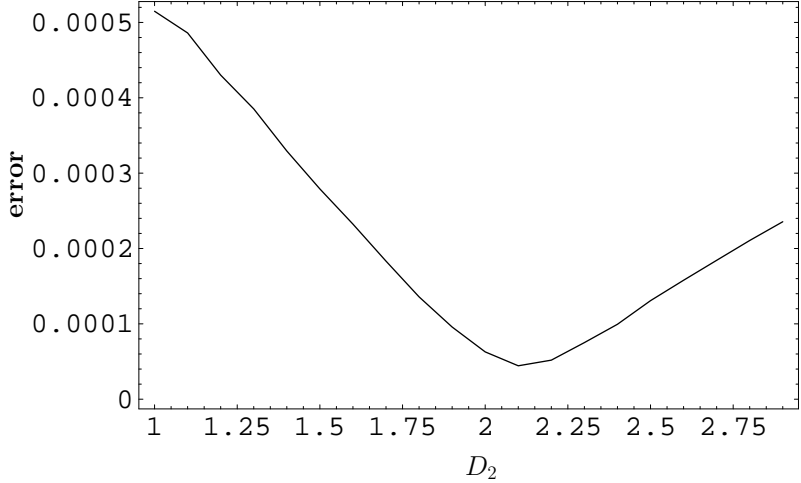}}}}
\vspace{-4.5cm}
\caption[phased]{{\small Extracting the scaling dimension $D_2$ by finite-size
scaling, in order to determine the kinetic term in
the effective action of the scale
factor.}}
\label{errd2}
\end{figure}
\begin{figure}[ht]
\vspace{-3cm}
\psfrag{x}{\bf{\Large $z$}}
\psfrag{T}{\Large\bf $P_{N_3}(z)$}
\centerline{\scalebox{0.6}{\rotatebox{0}{\includegraphics{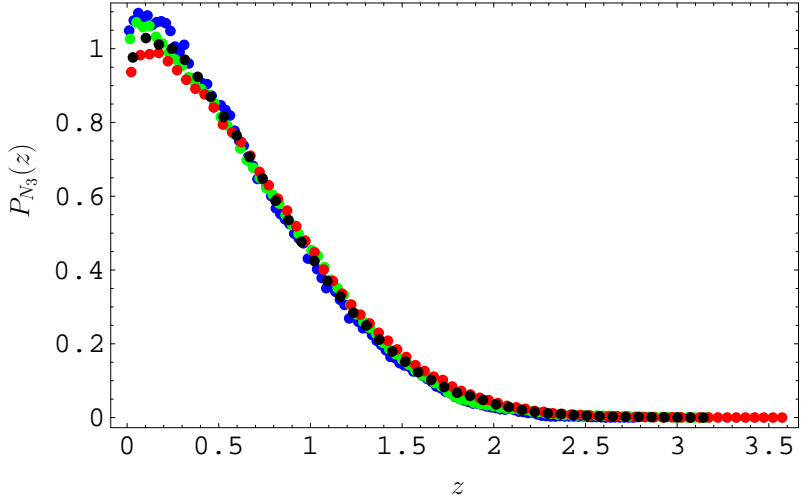}}}}
\vspace{-4.5cm}
\caption[phased]{{\small The distribution $P_{N_3}(z)$ of rescaled
volume differences $z$ between
adjacent spatial slices, for three-volumes $N_3\equ $ 20, 40, 80 and 160k.
}}
\label{VV2}
\end{figure}

Our first step is to determine the kinetic term in the effective action of
the dynamically triangulated model, and compare it to the term
$-a\dot a^2$ of the mini\-superspace model \rf{miniaction}. We obtain
this term by measuring the distribution of differences of adjacent
spatial volumes
\beq
z= \frac{|N_3(\tau+1)-VN_3(\tau)|}{N_3^{1/D_2}},~~~~N_3=N_3(\tau)+N_3(\tau+1).
\label{zdis}
\eeq
for different values of the discrete three-volume $N_3$. 
Again, the information we extract is two-fold:
we first determine the exponent $D_2$ from best overlap, and subsequently
the overall shape of the universal distribution.
In order to maximize the sampled
three-volumes, we employ the free boundary conditions described in 
Sec.\ \ref{num} above.
That is, all configurations extend over three time steps,
$\Delta\tau\equ 3$, with only
a single vertex each at the initial time $\tau\equ 0$
and the final time $\tau\equ 3$,
and the relevant volume difference given by \rf{zdis}, with $\tau\equ 1$.
From taking measurements at system volumes $\tilde N_4\equ$ 10, 20, 40, 80
and 160k, we extracted a best value for $D_2$ by minimizing the failure of the
curves to overlap for different volumes, using the same method
described earlier
for the determination of the scaling dimension $D_H$.
The result is illustrated by the curve of Fig.\ \ref{errd2}
which has its minimum at about 2.12, which we interpret as evidence
for $D_2\equ 2$. The overlapping curves for the volume
distributions $P_{N_3}(z)$
at $D_2\equ 2$ are depicted in Fig.\ \ref{VV2}. Apart from $z$-values near zero,
the overlap is clearly good. Moreover, the curve is fitted
very well by a Gaussian
$e^{- c z^2}$, with a positive constant $c$ independent of $N_3$.

From estimating the entropy of spatial geometries, that is,
the number ${\cal N}(N_3)$ of three-dimensional configurations at
given three-volume $N_3$,
one expects \cite{abkv} corrections of
the form $-c' N_3^\a$, with $0\leq \a < 1$,
$c'>0$, to the exponent $-c \, z^2\equiv -c\, (\Delta N_3/\Delta\tau)^2/N_3$
in the distribution $P_{N_3}(z)$.
This leads us to the following form for the Euclidean
effective action, {\it valid
for large three-volume} $N_3(\tau)$,
\beq
S^{\rm eff}_{N_4} \propto \sum_{\tau=0}^t 
\left(\frac{c_1}{N_3(\tau)} \left(\frac{\Delta N_3(\tau)}{\Delta\tau}\right)^2
+c_2 N_3^{\a}(\tau) - \lambda N_3(\tau)\right),
\label{seff}
\eeq
with $0\leq \a <1$, and $c_1$, $c_2>0$. In \rf{seff},
$\lambda$ is a Lagrange multiplier to be determined
such that
\beq
\sum_{\tau=0}^t  N_3(\tau) = N_4.
\label{lagrange}
\eeq
Unfortunately it is impossible to measure the precise form of the
correction term $c_2 N_3^\a$ directly in a reliable way.
However, it is clear from a
general scaling of the above action that
the only possibility for obtaining the observed scaling law,
expressed in terms of
the variable $\tau/N_4^{1/4}$, is by setting $\a\equ  1/3$.\footnote{In 
order to correctly reproduce the dynamics
for small three-volume, namely,
the stalk observed at large times $\tau$, the function
$N_3^{1/3}$ in \rf{seff} with $\a\equ 1/3$ must be replaced by a function of $N_3$
whose derivative at 0 goes like
$N_3^\nu$, $\nu\geq 0$, as explained in \cite{semi}. A simple choice is the
replacement $N_3^{1/3} \to (1+N_3)^{1/3}-1$.}

\begin{figure}[ht]
\vspace{-3cm}
\psfrag{d}{\bf{\Large $x$}}
\psfrag{VV}{\Large\bf $c_{\tilde N_4}(x)$}
\centerline{\scalebox{0.7}{\rotatebox{0}{\includegraphics{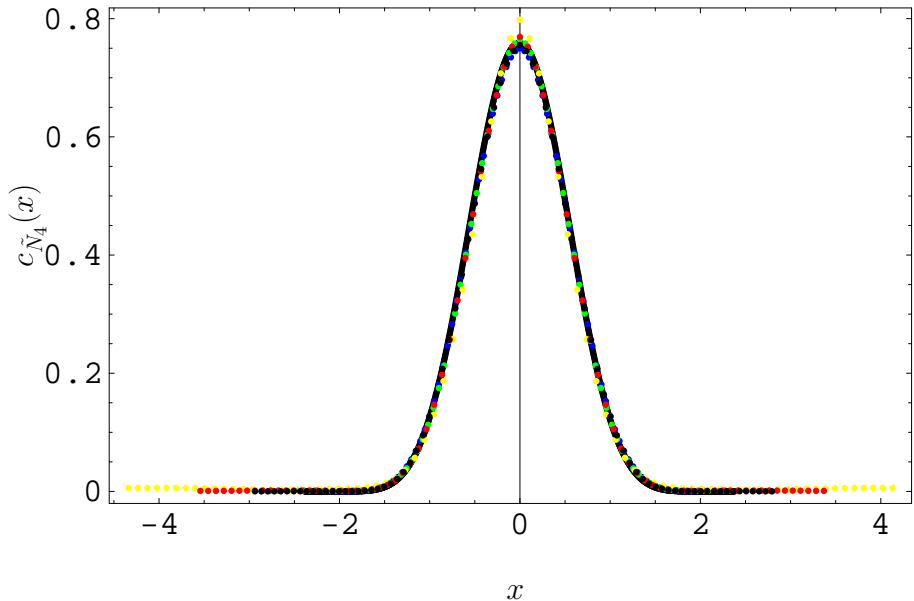}}}}
\vspace{-4.5cm}
\caption[phased]{{\small The scaling of the volume-volume distribution,
as function of the rescaled time variable $x\equ \delta/(\tilde N_4^{\it eff})^{1/4}$,
taken from Fig.\ \ref{scscs1}, but now superimposed with the correlator
obtained when using the minisuperspace action \rf{fullact} to generate a
distribution of three-geometries (black curve).
}}
\label{scscs2}
\end{figure}

Reverting back to a formulation in terms of the scale factor $a(\tau)$,
and by suitable rescaling of $\tau$ and $a(\tau)$), we can now write the
effective action of causal dynamical triangulations at fixed continuum four-volume
$V_4$ as
\beq
S^{\rm eff}_{V_4} = \frac{1}{G} \int_0^t d\tau \;
\left( a(\tau)\left(\frac{d a(\tau)}{d\tau}\right)^2
+ a(\tau) - \lambda a^3 (\tau)\right).
\label{fullact}
\eeq
Again, this expression is valid for sufficiently large scale factor,
since we have
explicitly dropped any reference to the quantum corrections at small $a$.
Comparing now with \rf{miniaction}, the startling conclusion is that we have
indeed rederived the Euclidean
minisuperspace action. However, here we have done it from first principles,
and {\it up to an overall sign}.
The strongest evidence so far that the effective action is indeed given
by \rf{fullact} comes from using it as {\it input} to create an artificial
distribution of three-volumes $N_3(\tau)$ \cite{semi}. We then evaluate 
the volume-volume correlator $c_{\tilde N_4}(x)$ of eq.\ \rf{volvolre} on this 
ensemble in the same way as we would do for genuine Monte Carlo data. 
As illustrated in Fig.\ \ref{scscs2}, the matching of this new curve with the previous 
Monte Carlo results (c.f. Fig.\ \ref{scscs1}) is impressive.

One immediate conclusion is that the collective effect of the dynamical
degrees of freedom other than $a(\tau)$, which are ignored in
minisuperspace models, is to change the sign of the kinetic term for the
scale factor (the ``global conformal mode") from $-a\dot a^2$ to
$+a\dot a^2$. This is the same nonperturbative cure of the conformal
divergence of the Euclidean path integral already encountered
in dimension 3 \cite{ajl3d,adjl} and argued for from a continuum
point of view in \cite{loll}.
It provides further evidence that the
well-known sicknesses and ambiguities of the Euclidean path integral
approach are features of the semiclassical and symmetry-reduced
approximations, not shared by a full nonperturbative treatment which is
rooted in a causal, Lorentzian formulation.

We have begun an exploration of the consequences of our result for
cosmology and quantum cosmology. When searching for a physical
interpretation, one must keep in mind that the effective action \rf{fullact}
we extracted directly from the computer simulations holds for
fixed four-volume. Translating between propagators of fixed four-volume
and fixed cosmological constant requires a further Laplace transformation
\cite{semi}.
Moreover, we are still in the Euclidean sector of the theory, and must
Wick-rotate back to obtain physical, Lorentzian results.

Nevertheless, it is exciting to see that the computation of some physical
quantities is already within reach: for example, we showed in
\cite{semi} how to derive the semiclassical wave function of the
universe without explicitly performing an inverse Wick rotation. 
The wave function can be compared directly with those arising in 
standard minisuperspace
treatments, like those of Vilenkin \cite{vilenkin,vilrev},
Hawking \cite{hh,eqg} and Linde \cite{linde}.
To relate to the dynamics of the real universe,
especially at very early times,
we will need to couple matter to our causal geometries. These
issues are currently being investigated and will be reported in
the near future.

\section{Nonclassical structure of the universe:\\
The geometry of spatial slices}\label{thin}

Having established that our quantum universe possesses certain classical
features at sufficiently large scales, we are interested in a more
detailed characterization of its geometry. We report in this section on various
measurements associated with the geometry of spatial slices
$\tau\equ const$, and in the next section on that 
of ``thick spatial slices", by which we mean
geometries of time extension $\Delta\tau \equ 1$, which are
``sandwiched" between adjacent spatial
slices of integer time $\tau$. Our measurements describe certain
invariant geometric properties of the ground state of spacetime,
but we make at this stage no attempt to relate them to actual observables
that one could go out and measure in the real universe. They will merely
serve to illustrate the complexity of the geometric structure and
highlight some of its nonclassical features.

Because our universes have a well-defined global notion of (proper)
time $\tau$, it is relatively straightforward to perform measurements within
a slice of constant $\tau$. We first restrict ourselves to
slices $\cal S$ of integer-$\tau$. Such slices 
consist entirely of spatial tetrahedra (whose edges
are all spacelike). The slices are of the form of spatial interfaces of
topology $S^3$ between adjacent sandwiches made out of four-simplices.  
As we will see, the Hausdorff dimension of the slices turns out to be three, 
as one would have naively expected in a four-dimensional universe. However,
the slice does not behave three-dimensionally under diffusion, 
and we will see that its structure is fractal in 
a precise sense which will be described below.

\subsection{Hausdorff dimension}

Our algorithm for determining the Hausdorff dimension $d_h$ of the spatial
slices is as follows. From a given $\cal S$ of discrete volume $N_3$,
we randomly pick a tetrahedron $i_0$. From $i_0$, we move out by
one step and obtain $n(1)$ tetrahedra at distance 1 from $i_0$ 
(that is, its nearest neighbours). Moving out by a further step, 
there will be $n(2)$ tetrahedra
at distance 2, and so on until all tetrahedra have been visited.
The numbers $n(r)$ recorded for each distance $r$ sum up to   
\beq
\sum_{r\equ 0}^{r_{max}} n(r)=N_3.
\label{v3av}
\eeq
Finally, we measure the average linear extension
\beq
\langle r \rangle = \frac{1}{N_3}\sum_r r n(r)
\label{extenav}
\eeq
of the spatial slice. 
For a slice of volume $N_3$, this process is repeated $N_3/50 +1$ times 
with different randomly chosen initial tetrahedra. The expectation value
$\langle r \rangle$ is then averaged over these
measurements. In this way, we obtain for every slice a data pair
\beq
\{ \langle\langle r \rangle\rangle, N_3\},
\label{pair}
\eeq
representing one ``bare" measurement. 

\begin{figure}[ht]
\vspace{-3cm}
\psfrag{log2}{\bf{\Large $\log(\langle r\rangle + 1.75)$}}
\psfrag{log1}{\Large\bf $\log(\langle N_3\rangle)$}
\centerline{\scalebox{0.6}{\rotatebox{0}{\includegraphics{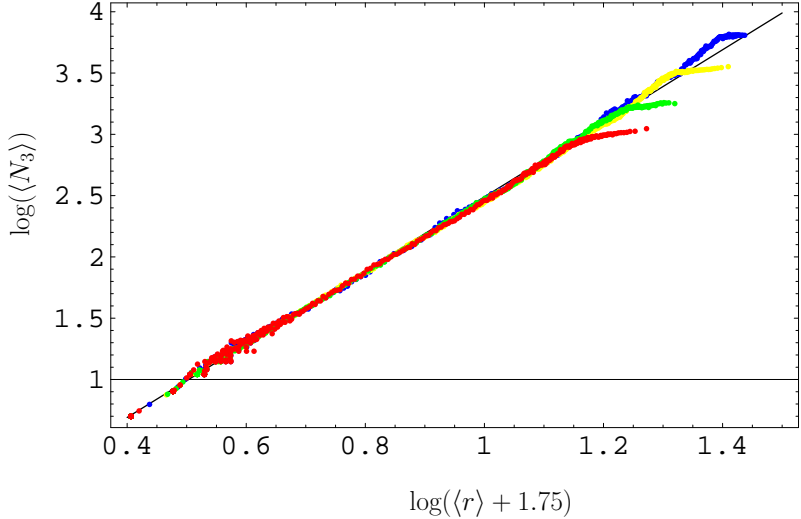}}}}
\vspace{-4.0cm}
\caption[phased]{{\small Log-log plot
of the average linear geodesic size $\langle r\rangle_{N_3}$ versus the
three-volume $N_3$, measured for $\Delta\equ 0.4$.
The straight line corresponds to a Hausdorff
dimension $d_h\equ 3$. Similar measurements for $\Delta\equ 0.5$ and
$0.6$ yield virtually indistiguishable results.
}}
\label{haus2.2b4}
\end{figure}

This process is performed for all $t$ spatial slices of the geometry, 
thus collecting $t$ data pairs.
We have typically done this for about 1000 different spacetime
geometries, obtaining altogether $1000t$ measurement points.
The final results are sorted by their $ \langle\langle r \rangle\rangle$-value
(in practice a continuous variable) 
and averaged over a sequence of 100 consecutive
points. This reduces the number of data points to $10t$, which are 
then displayed on a log-log plot. In the presence of finite-size scaling,
we expect them to follow a curve
\beq
\langle N_3\rangle (r) \propto (\langle r\rangle +r_0)^{d_h},
\label{3dscal}
\eeq
defining a spatial Hausdorff dimension $d_h$. 
In relation \rf{3dscal},
we have included a finite shift $r_0$ of the linear size $r$, based on earlier 
experience with finite-size corrections in the scaling relations \cite{ajw,aa}.
Our results, which show universal behaviour and scaling,
are presented in Fig.\ \ref{haus2.2b4}.
The shift value $r_0\equ 1.75$ gives the best linear fit. The resulting value of 
$d_h \approx 3$ is practically independent of $r_0$.

\subsection{Spectral dimension}

Given the results of the previous sections,
one might be tempted to conclude that the geometry of our dynamically
generated ground state simply {\it is}
that of a smooth four-dimensional classical
spacetime, up to Gaussian fluctuations. A more detailed analysis of the
geometry of spatial slices makes explicit that this is not so. -- We start by
measuring the spectral dimension $d_s$ of the spatial slices by
studying diffusion on them. Analogous to the
four-dimensional case described in Sec.\ \ref{fourspectral},
we choose an initial probability distribution 
\beq
K_T(i,i_0;\sg\equ 0) = \delta_{i,i_0}
\eeq
on the three-dimensional
triangulation $T$ of the slice $\cal S$, which is
zero everywhere except at a randomly chosen tetrahedron $i_0$,
and define the evolution in discrete time $\sigma$ by the rule
\beq
K_T(j,i_0;\sg+1) = \frac{1}{4}\sum_{k\to j} K_T(k,i_0;\sg),
\eeq
where $k\to j$ labels the (four) nearest neighbours of the tetrahedron $j$.
We measure the return probability $P_T(i_0;\sg)\equ K_T(i_0,i_0;\sg)$.
The measurements of the return probabilities are repeated a number
of times for each slice $\cal S$ with different starting
points $i_0$, precisely as in the measurement
of the Hausdorff dimension of $\cal S$. They are
then repeated for independent configurations (four-dimensional
triangulations generated by the Monte Carlo-procedure), again with the
same statistics as in the measurement of the Hausdorff dimension
of the three-dimensional spatial slices. Results
are stored for averaged slices with volumes $N_3\equ  n\times 500 \pm 50$,
because we expect them to depend on the system volume.
In this way we define the average return probability $P_{N_3}(\sg)$ and
extract the spectral dimension 
\beq
d_s(\sg) = -2 \frac{d\log P_{N_3}(\sg)}{d\log \sg}
\eeq
from the logarithmic derivative
exactly as in the four-dimensional diffusion process.

\begin{figure}[ht]
\vspace{-3cm}
\psfrag{time}{{\Large $\sg$}}
\psfrag{Ds}{{\Large $d_s$}}
\centerline{\scalebox{0.6}{\rotatebox{0}{\includegraphics{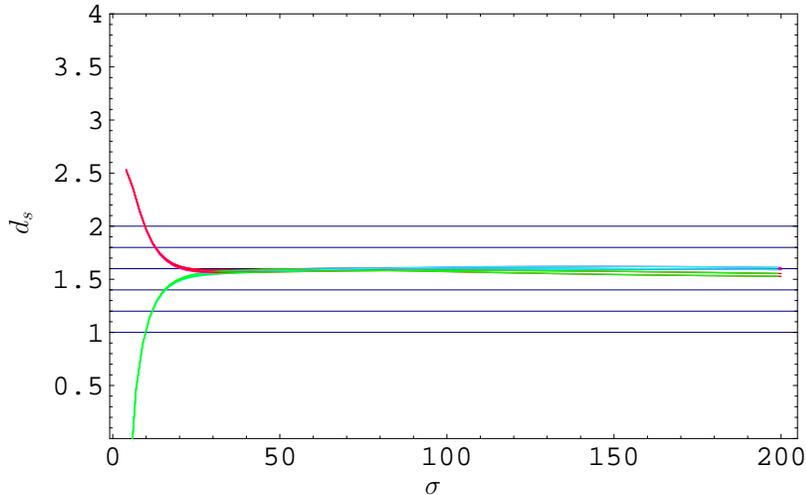}}}}
\vspace{-4.5cm}
\caption[phased]{{\small The logarithmic derivative of the return probability
for $\kappa_0\equ 2.2$ and $\Delta\equ 0.4$, and spatial volumes $N_3$ between
500 and 2000. The measurement
was performed for a system with $N_4\equ 91.1k$ and $t\equ 40$. The increasing
and decreasing lines correspond
to even and odd times, and illustrate the different behaviour for
short diffusion times $\sg$.}}
\label{ds2.2b4}
\end{figure}

The result is displayed in Fig.\ \ref{ds2.2b4} for even and odd times separately,
because of the different behaviour of $P_{N_3}(\sg)$ for short evolution times.
As can be seen the result is very stable for larger diffusion times
and gives $d_s \equ 1.56 \pm 0.1$ for
the spectral dimension, which is
only about half the expected ``classical" value!

\subsection{Critical exponents}\label{crit}

The noncanonical value of the spatial spectral dimension shows
that the quantum geometry has a nontrivial microstructure, which 
definitely is not that of a smooth four-dimensional manifold.
In fact, we already have come across further evidence of this fact 
in the measurement of the spacetime spectral dimension 
which approached the value four only at long distances. 
One way to
obtain an additional characterization of geometries at a fixed time 
is through the
functional form of their ``entropy" or ``number of states" ${\cal N}(N)$ as
function of their discrete volume $N$. Two different functional
forms for the entropy have appeared in previous studies of
simplicial {\it three}-dimensional quantum gravity \cite{abkv}, namely,
\bea\label{ja30} {\cal N}(N_3) &\propto&
N_3^{-3+\gamma_{str}}e^{\mu_0 N_3},\\
\label{ja31}
{\cal N}(N_3) &\propto&
e^{\mu_0 N_3 -c_1 N_3^{\nu}+ \cdots},\quad \nu < 1,
\eea
with a nonuniversal constant $\mu_0$ measuring the leading exponential
growth. The critical exponent $\gamma_{\rm str}$ in \rf{ja30} 
characterizes the fractal structure of the geometry \cite{jain,ajt,ajjt}.
The larger the value of $\gamma_{\rm str}$,
the more pronounced the tendency for a given three-geometry 
to develop so-called baby universes\footnote{In the present context, 
a ``baby universe" is a part of the
three-dimensional spatial triangulation 
which is connected to the remainder by a ``minimal neck''. This means
that the interface between the triangulated baby universe and the
``parent universe" is topologically a sphere and of minimal size.
Equivalently, it is given by
the surface of a tetrahedron consisting of four triangles. This is
easier to visualize in a two-dimensional simplicial manifold where 
the analogous minimal neck is an $S^1$ consisting of three edges.}.

The functional form \rf{ja31} on the other hand indicates a suppression 
of baby universes and formally may be seen as the limit
$\gamma_{\rm str} \to -\infty$ of \rf{ja30}. Below we will meet
both functional forms. It is important to understand that the 
``entropy'' ${\cal N}(N_3)$ in \rf{ja30} and \rf{ja31} does not merely
count the possible three-geometries, but also includes the weight 
coming from the action: in counting geometries we are making 
an importance sampling where
the action defines the probability weight of the geometry.
We saw above that the choice of
bare coupling constants in the four-dimensional theory
determines in which of the different phases of the lattice theory
we end up. The ``entropy'' of the spatial geometries at fixed time
is different in the three phases, even if the underlying geometries
are identical, and simply given by 
the space of all (abstract) triangulations of $S^3$. 

\begin{figure}[ht]
\vspace{-3cm}
\psfrag{x}{\bf{\Large $\log x$}}
\psfrag{px}{\Large\bf $\log P(x)$}
\centerline{\scalebox{0.6}{\rotatebox{0}{\includegraphics{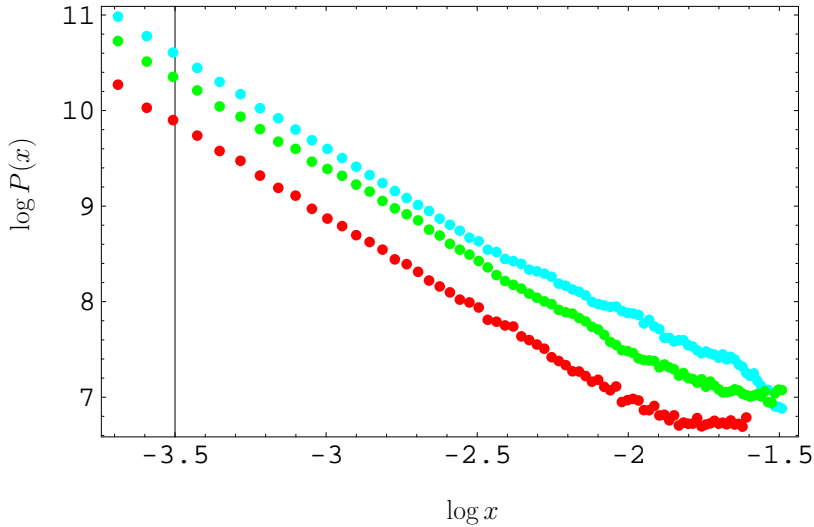}}}}
\vspace{-4.5cm}
\caption[phased]{{\small The distribution of the number of baby universes
with minimal neck as a function of their rescaled size $x$, for $\kappa_0\equ 2.2$
and $\Delta\equ 0.4$. Distributions are normalized by the total number of
baby universes in constant-time slices with $N_3 > 400$.
This number grows with the total spacetime volume $N_4$,
which is 45.5k (bottom), 91k (middle) and
181k (top curve).}}
\label{gamma2.2b4}
\end{figure}

If the functional form of the entropy is given by eq.\ \rf{ja30},
a convenient way of measuring the exponent $\gamma_{\rm str}$
is through the distribution of baby universes \cite{ajt,ajjt}.
Calling $B$ the volume of the baby universe (the number
of tetrahedra contained in it) and $N_3$
the total volume of the spatial geometry, we expect a distribution of
the form \cite{ajt,ajjt}
\beq
P(x) \propto x^{-2+\gamma_{\rm str}}\left(1+\frac{c(N_3)}{x} + \cdots\right),
\label{minbu}
\eeq
where
\beq
x=\frac{B}{N_3}\left(1 - \frac{B}{N_3}\right),\quad 0 < x <1/4,
\eeq
and $c(N_3) \propto 1/N_3$ is a volume-dependent constant.
Relation \rf{minbu} is expected to hold for sufficiently large $B$.
(It was derived under the assumption that
the samples of both baby and parent
universes have an asymptotic distribution of the form
\rf{ja30}.) Since it is very difficult in the present set-up to collect data
for any particular fixed value of $N_3$, we have performed
``integrated" measurements to determine $\gamma_{\rm str}$,
by collecting the distribution of the number of baby universes of a given
$x$  for various volumes $N_3 > 400$.
Also in this case we expect the distribution of baby universes to follow
\beq
P(x) \propto x^{-2+\gamma_{\rm str}}\left(1+\frac{d}{x}+ \cdots\right),
\label{minbuav}
\eeq
with a constant $d$ depending now on the range of three-volumes $N_3$ over
which we have averaged the distribution.
The resulting distributions for $\kappa_0\equ 2.2$, $\Delta\equ 0.4$
and total system sizes $N_4=$ 45.5, 91 and 181k
are shown in Fig.\ \ref{gamma2.2b4}.
Although the measurement is of rather poor quality, and
-- depending on which range of $x$ is used for the fit --
the uncertainty in $\gamma_{\rm str}$ rather big,
the approximate linearity of the plots strongly suggests
that the distribution is
indeed dominated by a power law for $x$.
Fitting $\gamma_{\rm str}$ to formula \rf{minbuav}, we obtain
$\gamma_{\rm str}\equ 0.35 \pm .09$.

Another, seemingly independent,
way of characterizing the short-distance behaviour of the spatial
slices $\cal S$ is by measuring the coordination number of vertices in
$\cal S$, i.e. the number $n_v$ of spatial tetrahedra sharing a given
vertex $v$.\footnote{Note
that for a 3d manifold there is a linear relation between the coordination
number of a vertex $v$ and the number of links emerging from $v$.}
In the simulations, we observe vertices with
relatively high coordination numbers $n$, and a
distribution of $n$ which can be approximated a simple power law
\beq
P(n) \propto \frac{1}{(n+n_0)^{1+\alpha}}.
\label{power}
\eeq
As usual, we have included a finite shift $n_0$ to take into
account finite-size effects.
For a three-dimensional triangulation, $n \ge 4$ and
$n$ may take only even values.

\begin{figure}[ht]
\vspace{-3cm}
\psfrag{x}{\bf{\Large $\log(n+2)$}}
\psfrag{y}{\Large\bf $\log P(n)$}
\centerline{\scalebox{0.6}{\rotatebox{0}{\includegraphics{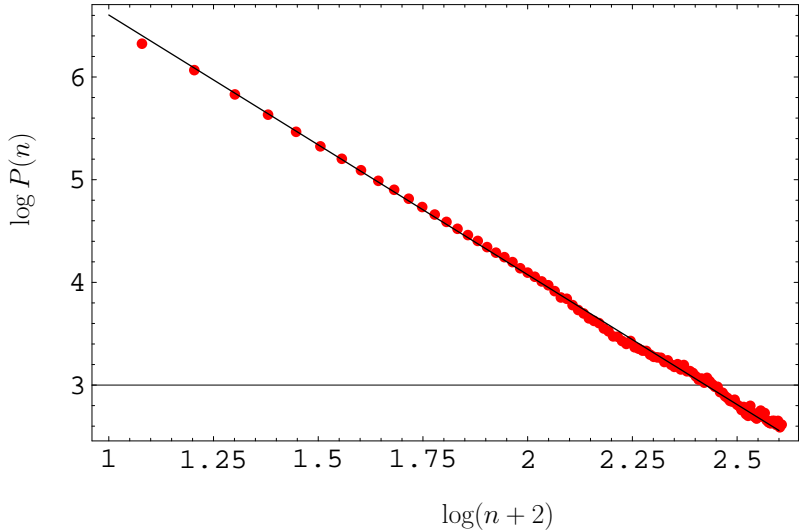}}}}
\vspace{-4.5cm}
\caption[phased]{{\small Distribution $P(n)$ of the coordination numbers $n$ of
vertices in a three-dimensional spatial slice, for $\kappa_0\equ 2.2$
and $\Delta\equ 0.4$.
The straight line represents a numerical fit to formula
\rf{power}, with $n_0\equ 2$.}}
\label{vcoord2.2b4}
\end{figure}

Fig.\ \ref{vcoord2.2b4} illustrates the distribution of coordination numbers for
$4 \le n \le 200$, measured for a
system with $\kappa_0\equ 2.2$, $\Delta\equ 0.4$, $t\equ 40$ and system size 
$N_{4}=$181k.
The data points have been collected from spatial slices with three-volume
$N_3 > 500$, and can be seen to behave perfectly
linearly on the log-log plot.
The straight line in Fig.\ \ref{vcoord2.2b4} corresponds to a fit with
$\alpha\equ 1.53$. 
In theories of random geometry, this type of
power-law behaviour of the vertex coordination number 
was first observed in dynamically triangulated bosonic string
theory \cite{adf}. 
The value of the exponent
$\alpha$ extracted from the data by $\chi^2$-fits
depends slightly on the choice of $n_0$ and of the cut-off in the data.
Our estimate is $\alpha\equ 1.53\pm 0.05$.

\begin{figure}[ht]
\vspace{-3cm}
\psfrag{x}{\bf{\Large $\log N_3$}}
\psfrag{y}{\Large\bf $\log\langle n_{max}\rangle$}
\centerline{\scalebox{0.6}{\rotatebox{0}{\includegraphics{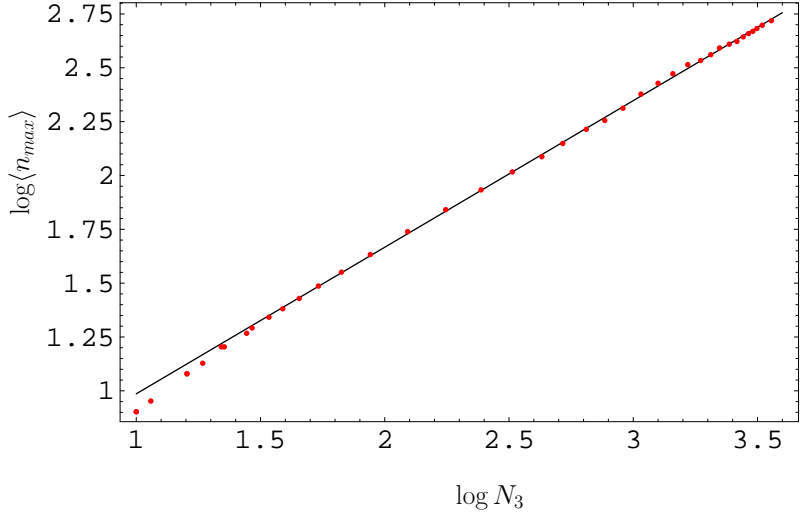}}}}
\vspace{-4.5cm}
\caption[phased]{{\small Distribution of the largest
coordination number $n_{max}$ as
a function of the spatial volume $N_3$,
for $\kappa_0\equ 2.2$, $\Delta\equ 0.4$, $t\equ 40$ and
$N_4\equ 91.1k$.}}
\label{top2.2b4}
\end{figure}

Another way to determine $\alpha$ is by studying the
distribution of points with
the largest coordination number as a function of the
spatial volume $N_3$ (or the total number
of vertices $N_0(N_3)$, the two numbers being proportional for the
phase of four-dimensional gravity we are considering here).
One expects a behaviour of the form
\beq
P(n > n_{max}(T)) \propto \frac{1}{N_3(T)},
\label{nmax1}
\eeq
where $n_{max}(T)$ denotes the maximal coordination number
for a particular three-dimensional triangulation $T$. 
The proportionality \rf{nmax1} is expected
to be valid if the distribution $P(n)$ is such that a
typical $n_{max}(T)$ is so small that it is not influenced
by the fact that the volume $N_3(T)$ contains only a finite number
$N_0(T)$ of vertices. (Clearly, $n_{max}(T)$ cannot be larger
than $N_0(T)/2$.)
Under this assumption the orders $n_v$ of the $N_0(T)$ vertices $v$
in $T$ are representative of the distribution $P(n)$. Since there
are $N_0(T)$ of them, the chance of ``drawing'' an (abstract) vertex
from the distribution $P(n)$ with a coordination number 
$n > n_{max}(T)$ is of the
order $1/N_0(T)$, which, given $N_0(N_3) \propto N_3$, is
equivalent to \rf{nmax1}.
Consequently, using the scaling \rf{power}, we obtain that
\beq
n_{max}(N_3) \propto N_3^{1/\alpha}.
\label{nmax2}
\eeq
To check this property, we measure for each surface $\cal S$
the largest coordination numbers in $\cal S$.
Since these fluctuate considerably, we measure the average of the three
largest values for each $\cal S$. We then follow a strategy similar
to that adopted for the measurement of the Hausdorff dimension.
Namely, we collect data pairs $\{N_3,n_{max}\}$,
sort the resulting set by volume $N_3$ and average over a set of
1000 measurements. Fig.\ \ref{top2.2b4} shows the
result in the form of a log-log plot, confirming the expected linear behaviour.
The solid line corresponds to the value
\beq
\frac{1}{\alpha} = 0.681,
\eeq
in fair agreement with the previous
measurement from smaller correlation numbers,
and the fit gives $1/\alpha\equ 0.681 \pm .05$, translating roughly into
$\alpha\equ 1.468\pm 0.12$. The origin of the uncertainty is the chosen range in
the three-volume $N_3$ taken into account in the fitting.

\subsection{Relation to branched polymers}

The above measurements lead to the
intriguing conclusion that {\it all} the
geometric properties of the spatial slices
measured so far can be modelled by a particular kind of
branched polymers\footnote{In combinatorics, branched polymers are 
conventionally called rooted trees, where ``rooted'' means
that one link or vertex is marked, and therefore distinguished from the rest.},
as we will now go on to explain.

A (planar) branched polymer is an (abstract)
one-dimensional polymer which admits
arbitrary branching. It is modelled by (abstract) connected planar graphs
which contain no loops. As a consequence, 
cutting any link of such a graph will result in two
disconnected graphs. At each vertex an ``incoming'' link is
allowed to ``branch'' into $n\mi 1$ ``outgoing'' links with
probability $p_n$. The statistical model of (rooted)
branched polymers is then defined by the partition function
\bea\label{ja50}
Z_{B\! P}(\mu) &=& \sum_N e^{-\mu N} \; Z_{B\! P}(N)\\
Z_{B\! P}(N)&=& \sum_{B\! P_N} \rho(B\! P_N) , \quad 
\rho(B\! P_N)=\prod_{n\equ 1}^N p_n,
\label{ja51}\eea
where the summation is over all branched polymers with $N$ vertices,
and the weight $\rho(B\! P)$ is the product
of the branching probabilities. One can also embed the abstract
branched polymers in a target space, either a lattice or $R^d$.
In the latter case, it is convenient to work with a quadratic action $(x_u-x_v)^2$
for pairs of nearest-neighbour vertices $u$ and $v$ with coordinates $x_u$ and
$x_v$, and perform an additional
integration over the target-space coordinates in the partition function.

The fractal properties of branched polymers are well understood.
The leading behaviour for large $N$ is an analogue\footnote{The
difference between the power $\gamma\mi 2$ in eq.\ \rf{ja52}
and $\gamma \mi 3$ of eq.\ \rf{ja30} arises because of the use of
{\it rooted} branched polymers, which gives rise to an extra
factor of $N$.} of eq.\ \rf{ja30}, namely,
\beq\label{ja52}
Z_{B\! P}(N) \sim N^{\gamma -2} \; e^{\mu_0 N},
\eeq
where $\gamma$ can take a range of real values, depending on 
the branching probability $p_n$.
The intrinsic Hausdorff dimension $d_h^{(in)}$ of
branched polymers with $\gamma > 0$ is \cite{adj,adj1,jk}
\beq\label{ja53}
d_h^{(in)}= \frac{1}{\gamma}, 
\eeq
with spectral dimension
given by \cite{thordur-john,anrw}
\beq
\label{ja54}
d_s= \frac{2}{1+\gamma}.
\eeq

We note that formulas \rf{ja53} and \rf{ja54} already provide a close
match with our data if we identify $\gamma$ with the $\gamma_{\rm str}$
of the simulations. 
As we will see, there is indeed a concrete branched-polymer model 
which has the $\gamma$-value we have observed.
In regularized bosonic string theory one is naturally
led to studying branched polymers with branching weight \cite{adf}
\beq\label{ja55}
p_n \sim \frac{1}{n^\alpha}~~~~{\rm for}~~n\gg 1.
\eeq
Generically, the value of $\gamma$ for a given $\alpha$ is
\beq\label{ja56}
\begin{array}{llll}
\gamma(\alpha) &=& 1/2 &{\rm for} \quad \alpha \leq 1, \\
\gamma(\alpha) &=& 1-\alpha &{\rm for} \quad 1 < \alpha,
\end{array}
\eeq
which does not cover the $\gamma$-range we are interested in.
However, if the partition function satisfies a certain technical
condition\footnote{For \rf{ja58} to hold one needs
\beq\label{ja60}
\frac{d Z_{B\! P}(\mu)}{d\mu} < 0\quad {\rm for}\quad \mu > \mu_0,
\quad\quad\frac{d Z_{B\! P}(\mu)}{d\mu}\Big|_{\mu\equ \mu_0}= \infty,
\eeq
which is naturally satisfied for some
branched polymer models coming from noncritical string
theory \cite{adj1}.}, the relations \rf{ja56} are changed to 
\beq\label{ja58}
\begin{array}{llll}\gamma(\alpha) &=&
{(\alpha-1)/{\alpha}} & {\rm for}\quad  1< \alpha <2\\
 \gamma(\alpha)& =& {1/2}& {\rm for} \quad 2\leq \alpha
\end{array}\eeq
(\cite{jk}, and see \cite{book} for
a detailed discussion).
While the associated condition \rf{ja60} seems rather artificial from
the perspective of branched polymers,
it appears in a natural way if the branched polymers emerge as
a limit of higher-dimensional models like noncritical strings
with central charge $c > 1$. As a matter of fact, the generic value
of $\gamma$ in this case is 1/3 \cite{adj1}. 
Summarizing the critical exponents for this particular class of branched 
polymers,
\beq\label{ja59}
\gamma=\frac{1}{3},\quad d_h^{(in)}=3,\quad d_s=\frac{3}{2},
\quad \alpha=\frac{3}{2},
\eeq
we see that within the measuring accuracy, {\it they are
precisely the numbers characterizing the spatial geometries of
causal dynamical triangulations at constant time}. 
Although the vertex order $n_v$ of our three-dimensional
triangulations has {\it a priori} nothing to do with the branching
of polymers, the above results are a strong hint that there
effectively is such a connection. While this
confirms our earlier assertion that the geometry of
the spatial slices behaves very nonclassically, it opens the
exciting possibility that quantum gravity as described by 
this model may be amenable to analytic approximations,
precisely as was the case in the study of noncritical strings
with $c > 1$ \cite{adj1}.

\section{Nonclassical structure of the universe:\\
The geometry of ``thick" spatial slices}\label{thick}

Another substructure of spacetime whose geometry can be investigated
straightforwardly are what we shall call ``thick" spatial slices, namely, the
sandwiches of geometry contained in between two adjacent spatial slices
of integer times $\tau$ and $\tau+1$, with topology $I\times S^3$.
As illustrated in Fig.\ \ref{4dsimplices}, they are made up
of four-simplices of types (4,1) and (3,2) and their time-reversed counterparts
(1,4) and (2,3). Because of their finite time extension, one might expect the
geometry of the thick slices to behave more ``classically", and this is indeed
corroborated by our simulations, which include a measurement of the
Hausdorff and spectral dimension of the thick slices.

A convenient way to represent a sandwich geometry is by considering
the three-dimensional piecewise flat geometry that results by cutting
the thick slice at time $\tau+1/2$. The analogous procedure in one
dimension less, which is easier to visualize, gives a generalized
two-dimensional triangulation consisting of triangles and squares
\cite{ajl3d,ajlv,ajlabab}.
The triangles are the intersection surfaces of (3,1)- and
(1,3)-tetrahedra, and the squares those of (2,2)-tetrahedra. In one
dimension higher, the fundamental building blocks are no longer
tetrahedra but four-simplices. Consequently, the intersection patterns
are three-dimensional flat building blocks, namely, spatial tetrahedra
from slicing the (4,1)- and (1,4)-simplices, and spatial triangular
prisms from slicing the (3,2)- and (2,3)-simplices.  In order to
distinguish between a tetrahedron that comes from cutting a
(4,1)-simplex from one associated with an upside-down, (1,4)-simplex,
one colour-codes the links of the intersection hypersurface into red 
and blue, say. The red links are those that will be ``shrunk away"  
when the constant-time surface approaches the integer-time $\tau\pl 1$,
and similarly the blue links will disappear as the hypersurface 
is moved towards time $\tau$. As a result, the tetrahedra have all links of one
colour, whereas the prisms are always bi-coloured (see also \cite{blackhole}).
Note that the average spatial volume in terms of
counting three-dimensional building blocks is larger (by a factor of
about 2.5) at slices of half-integer time, because -- unlike at
integer times -- all four-simplices between $\tau$ and $\tau+1$
contribute to the counting.

\subsection{The Hausdorff dimension}

\begin{figure}[ht]
\vspace{-3cm}
\psfrag{log1}{\bf{\Large $\log(\langle r \rangle +2)$}}
\psfrag{log2}{\Large\bf $\log\langle N\rangle$}
\centerline{\scalebox{0.6}{\rotatebox{0}{\includegraphics{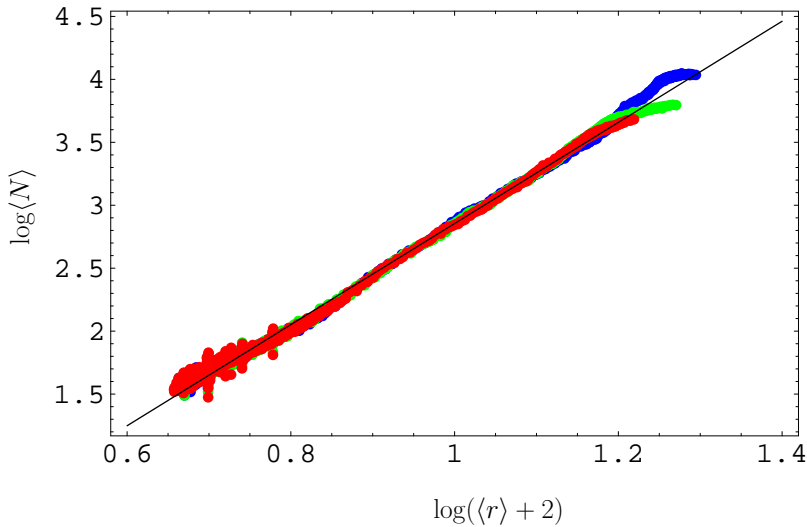}}}}
\vspace{-4.5cm}
\caption[phased]{{\small Log-log plot
of the average linear geodesic size $\langle r\rangle_{N}$ versus the
volume $N$ of a thick slice. The straight line corresponds to a Hausdorff
dimension $d_H\equ 4$.
}}
\label{Haus2.2b6.ps}
\end{figure}

To determine the Hausdorff dimension $d_H$ of the thick slices,
we pick a random
four-simplex from such a slice, or, equivalently, a random three-dimensional
building block from the corresponding intersection pattern at
half-integer time. Continuing in this three-dimensional representation of
the slice geometry, we proceed exactly as we did when measuring the
Hausdorff dimension $d_h$ of the spatial slices at integer-$\tau$.
Starting from the
randomly chosen initial three-dimensional building block, we move out,
one step at a time, to the nearest neighbours at distance $r$, from the
previous set of building blocks at distance $r-1$. In terms of four-simplices,
this means we never move out to simplices earlier than $\tau$ or later
than $\tau \pl 1$, that is, by definition we never leave the thick slice.
At each step, we keep track of the number $n(r)$ of building blocks,
whose sum is equal to the total slice volume $N$,
\beq
\sum_{r\equ 0}^{r_{max}} n(r) = N,
\label{vav}
\eeq
and which enables us to compute the average linear size
\beq
\langle r \rangle = \frac{1}{N}\sum_r r n(r)
\eeq
of the thick slice. For a given slice of volume $N$,
this process is then repeated
$N_4/100+1$ times with randomly chosen initial building blocks. The average
size $\langle r\rangle$ is again averaged over this set and the data pair
$\{\langle\langle r \rangle\rangle, N\}$ stored. Using the same
procedure as described following formula (\ref{pair}) above, we extract the
Hausdorff dimension $d_H$ from the relation
\beq
N_4 \propto (\langle r \rangle + r_0)^{d_H},
\label{hausagain}
\eeq
with a finite constant $r_0$ taking into account finite-size corrections.
The results are presented in Fig.\ \ref{Haus2.2b6.ps} in the form of a
log-log plot.
The straight line corresponds to $d_H\equ 4$, and the best fitted value is
$d_H \equ 4.01\pm 0.05$. Measurements were
performed on spacetimes with total size $\tilde N_4=$20, 40 and 80k, at
$\kappa_0\equ 2.2$, $\Delta\equ 0.6$ and $t\equ 80$. -- We conclude that the
transition from a genuine constant-time slice to a thick slice adds one
extra dimension to the Hausdorff dimension -- the
thickened slice already ``feels''
some aspects of the four-dimensionality of the full spacetime.
Note that this would not be true for a thick slice of a classical, regular
lattice, whose Hausdorff dimension would be the same as that of a
``thin" slice. The likely mechanism for the dimensional increase in
the triangulations is the appearance of ``short-cuts" between two
tetrahedral building blocks, say, once one moves slightly away
from a slice of integer-$\tau$. 

\subsection{The spectral dimension}

\begin{figure}[ht]
\vspace{-3cm}
\psfrag{x}{{\Large $\sg$}}
\psfrag{y}{\Large $d_S$}
\centerline{\scalebox{0.6}{\rotatebox{0}{\includegraphics{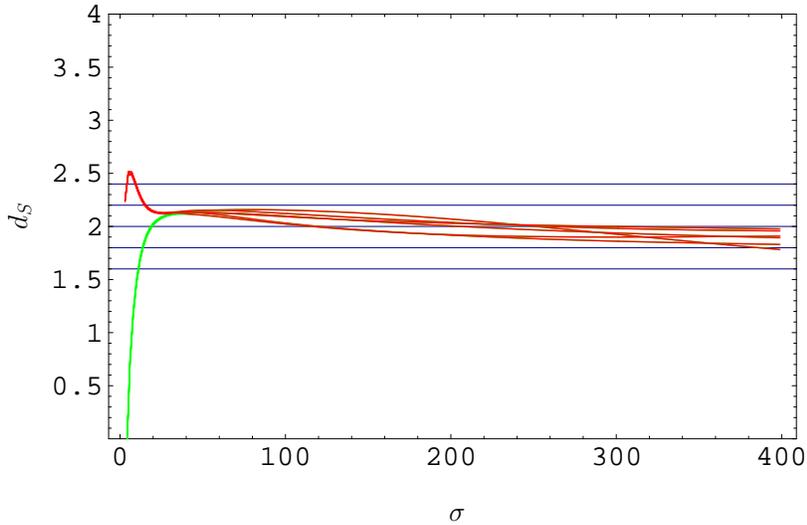}}}}
\vspace{-4.5cm}
\caption[phased]{{\small The logarithmic derivative of the return probability
for $\kappa_0\equ 2.2$ and $\Delta \equ 0.6$ and thick slices of volumes $N$
between 900 and 2000.
The measurement
was performed for a system with $N_4\equ 161k$ and $t\equ 80$.
Curves extrapolating between even evolution times decrease rapidly at
small $\sg$, whereas curves of odd times  stay above $d_S\equ 2$ at small times.
}}
\label{ds80}
\end{figure}

Next, we turn to the spectral dimension $d_S$ of
the thick slices. When adapting our earlier measurement of the spectral
dimension $d_s$ of spatial slices, we have to take into account
that the tetrahedral building blocks in the three-dimensional intersection
pattern at half-integer $\tau$ have four neighbours, whereas the prisms
have five.
We start from an initial probability distribution
\beq
K_T(i,i_0, \sg\equ 0)=\delta_{i, i_0}
\label{probthick}
\eeq
concentrated
at a randomly chosen three-dimensional building block with label $i_0$
in the three-dimensional ``triangulation'' $T$ (consisting of
tetrahedra and triangular prisms),
and define the evolution process by
\beq
K_T(j,i_0;\sg+1) = \sum_{k\to j} \frac{1}{g_k} K_T(k,i_0;\sg),
\eeq
where $g_i$ denotes the number of neighbours of building block $i$, and
$k\to j$ are the nearest neighbours of the ``simplex'' $j$ (which can
now also be a prism).
By construction, the total probability
\beq
\sum_i K_T(i,i_0;\sg) = 1
\eeq
is conserved. As before,
we measure the return probability $K_T(i_0,i_0;\sg)$. After
the appropriate averaging over initial points $i_0$ and triangulations $T$
we obtain the quantum return probability $P_N(\sg)$ as a function
of the number of building blocks $N$ in the triangulations $T$ (we
keep the number $N$ approximately constant in the triangulations
included in the average). The expected behaviour
\beq
P_N(\sg) \propto \sg^{-d_S/2}+\mbox{finite-size corrections}
\eeq
defines a spectral dimension $d_S$, which in general will be
different from the spectral dimension $d_s$ determined for the
constant-time slices.
The value of $d_S$ is extracted by measuring the logarithmic derivative
of $P_N(\sg)$ as a function of the evolution time $\sg$,
\beq
d_S(\sg)= -2 \;\frac{d\log P_N(\sg)}{d\log \sg}.
\label{thickevolve}
\eeq
The results of our measurements are displayed separately for even and odd
times in Fig.\ \ref{ds80}, since the behaviour of $P_N(\sg)$ is
different for very small $\sigma$ (as already observed in our previous measurements
of spectral dimensions).
As seen in the figure, beyond small $\sigma$ we do not find the perfect
constant logarithmic derivative observed earlier for
the ``thin" slices at integer time.
If we nevertheless treat the output curves as constants and take
their mean value as representative of the 
logarithmic derivative $d_S$ (assuming it {\it is} 
$\sg$-independent), we obtain
$d_S \equ  2.0 \pm 0.2$.
Although it lies above the value $d_s$ for the spectral dimension
of the spatial slices, this dimensionality is still highly nonclassical,
and reflects the presence of some intermediate regime between
a purely spatial and a purely spacetime geometry.

\subsection{Baby universes}

To further characterize the geometry of thick slices, we also looked into its
baby universe structure. Minimal necks can take the form of either four
triangles, forming the surface of a tetrahedron, or of two triangles and three
squares, forming the surface of a triangular prism (of course, {\it without}
an interior tetrahedron or prism being located there). It turns out that the
four-dimensional manifold constraints, which ensure that the four-simplices
are glued in a way that results in a simplicial {\it manifold}, impose severe
restrictions on the types of baby universes with minimal necks that can
occur.\footnote{Analogous manifold conditions for intersection graphs in
three-dimensional simplicial spacetimes were formulated in \cite{ajlv}.}
For a tetrahedral minimal neck, the baby universe which branches off
there can only contain building blocks which are tetrahedra of the same
colour. Minimal necks of the form of a prism cannot occur altogether.
We attempted to measure the distribution of ``monochrome" tetrahedral
baby universes, but found them to be small in size, within a range
practically independent of the slice volume $N$. We did not check for
baby universes with larger necks. Our tentative conclusion is that
the microscopic structure of the thick slices is more regular
than that of the genuine spatial slices of fixed constant time.

It is tempting to conjecture that the absence of baby universes
means that the entropy function ${\cal N}(V)$
for the thick spatial slices has the functional form \rf{ja31} rather
than \rf{ja30}, which was the form observed
for the thin spatial slices at fixed integer time. As mentioned in
Sec.\ \ref{crit}, the entropy behaviour \rf{ja31} signals an almost
complete suppression of baby universes. Moreoever, if the
exponent $\nu$ in eq.\ \rf{ja31} was equal to 1/3, it would fit
neatly with the observation that the effective macroscopic
action \rf{fullact}, which reproduces so well the macroscopic
four-dimensional features we observe in the simulations,
{\it does} contain the characteristic linear term in the scale factor $a(\tau)$
(corresponding to a factor $V_3^{1/3}(\tau)$ in terms of the continuum 
three-volume). 
It would also be in line with our expectation that the
structure of the thick spatial slices,
rather than that of the thin slices at fixed time should relate 
most directly to the global four-dimensional structure.

\section{Summary and outlook}\label{conclusion}

This paper describes the currently known geometric properties of the quantum
universe generated
by the method of causal dynamical triangulations, as
well as the general phase structure of the underlying statistical model of
four-dimensional random geometries.
The main results are as follows.
An extended quantum universe exists in one of the three observed phases
of the model,
which occurs for sufficiently large values of the bare Newton's constant $G$
and of the asymmetry $\Delta$, which quantifies the finite relative length
scale between
the time and spatial directions. In the two other observed phases, the
universe disintegrates into a rapid succession of spatial slices of vanishing
and nonvanishing spatial volume (small G), or collapses
in the time direction to a universe that only exists for an infinitesimal moment
in time (large G, vanishing or small $\Delta$). In either of these two cases,
no macroscopically extended spacetime geometry is obtained.

By measuring the (Euclidean) geometry of the dynamically generated
quantum spacetime\footnote{Since our universe is a weighted quantum 
superposition
of geometries, all ``measurements" refer to expectation values of
geometric operators in the quantum ground state.} 
in the remaining phase, in which the universe appears to be
extended in space and time, we collected strong evidence that it behaves as a
{\it four}-dimensional quantity on large scales. First, finite-size scaling of the
volume-volume correlator for spatial slices is observed when the time
variable is rescaled by the fourth root of the discrete four-volume,
eq.\rf{xdef}, within measuring accuracy. Second, the large-scale 
spectral dimension of the universe,
determined from a diffusion process on the quantum ensemble of
geometries, is compatible with the value 4.

The dynamical derivation of the correct classical result for the dimension
from a completely nonperturbative path-integral setting provides a
considerable boost to our efforts to construct a theory of quantum
gravity by the method of causal dynamical triangulations.
Furthermore, we were able to show that the universe possesses
other features of classical geometry at sufficiently large scales. Namely,
using the four-dimensionality at large scales, we derived an effective
Euclidean action for the scale factor of the universe (equivalently,
the volume of its spatial slices). For the case of fixed spacetime volume,
this action coincides {\it up to an overall sign} with a simple
minisuperspace action frequently used in quantum cosmology.
Again, we believe that this is a highly nontrivial and very exciting
result. It has already enabled us to derive a
``wave function of the universe" from first principles \cite{semi}. The
change in sign of the action implies that the Euclidean sector of
our quantum gravity model is well-defined. This is a welcome result since,
unlike in standard
quantum cosmology, no ad hoc cure is needed to deal with
a conformal kinetic term that renders the action unbounded below.
Some care is required in interpreting it, since we still need to
perform an inverse Wick rotation to get back to the physical,
Lorentzian sector of the theory. A detailed discussion will appear
elsewhere.

Returning to the analysis of the quantum geometry of our universe,
we gained further insights by performing measurements of a 
more local nature by looking at the intrinsic geometry
of ``thin" and ``thick" spatial slices. A thin slice is by definition a
spatial geometry at fixed integer time somewhere inside a universe,
and a thick slice a piece of spacetime contained between two
adjacent thin slices. We found that in general the measured
observables exhibit nonclassical behaviour, in the sense that
dynamically determined dimensionalities of the slices do not
coincide with the values one would have expected naively for
the corresponding submanifolds of a {\it classical} four-dimensional
spacetime. This result is not in contradiction with the four-dimensional
behaviour of the universe at large scales, but underscores the
fact that the local geometry of the quantum spacetime has
highly nonclassical features.\footnote{One should keep in mind that
slices of constant time are in no way physically distinguished
in a theory of pure gravity, and their geometric properties therefore
not directly related to physical observables.}

We found that the measured Hausdorff dimension
of thin slices comes out as 3 with good accuracy, but that the spectral
dimension is only around 1.5! By making additional measurements of 
the distributions of vertex
coordination numbers and of baby universes in the thin slices 
we were able to show that the entire set of parameters is matched
by the critical exponents of a particular class of branched polymers,
potentially opening a new window to an analytic investigation of quantum
spacetime.

The thick slices were found to have Hausdorff dimension 4 within measuring
accuracy, and a spectral dimension of around 2. This, and the
apparent absence of baby universes of appreciable size indicate
that the geometry of the thick slices represents a genuinely
intermediate case between that of the thin slices and of the
full spacetime. In particular, their geometry -- although far from being
classical -- seems to be more regular than that of the thin slices.

The most local measurement of quantum geometry so far is that
of the spectral dimension of spacetime {\it at short distances}, which
provides another quantitative measure of the nonclassicality of
geometry. As we have seen, the spectral dimension changes smoothly
from about 4 on large scales to about 2 on small scales. Not only
does this (to our knowledge) constitute the first dynamical 
derivation of a scale-dependent
dimension in full quantum gravity, but it may also provide a natural
short-distance cut-off by which the nonperturbative formulation
evades the ultra-violet infinities of perturbative quantum gravity.
Remarkably, evidence for a dimensional reduction from four to two has
been found recently in the so-called exact renormalization group
approach to (Euclidean) quantum gravity, truncated to a 
finite-dimensional parameter space \cite{lauscher}. One could be
particularly encouraged by the fact that despite very different starting 
points and quantization methods the two approaches  
nevertheless seem to lead to compatible results.

In summary, what emerges from our
formulation of nonperturbative quantum gravity as a
continuum limit of causal dynamical triangulations is a compelling 
and rather concrete geometric picture of {\it quantum spacetime}. 
Quantum spacetime possesses a number of 
large-scale properties expected of a four-dimensional classical
universe, but at the same time exhibits a nonclassical and nonsmooth
behaviour microscopically, due to large quantum fluctuations 
of the geometry at small scales. These fluctuations ``conspire" to
create a quantum geometry that is effectively two-dimensional
at short distances.

Of course, our large-scale measurements of geometry so far 
do not characterize the classical properties of our quantum spacetime
completely, and further work is under way to determine its
large-scale curvature properties, and check that
Newton's inverse square law can be recovered in an appropriate limit
(see \cite{newton} for related previous attempts).
Having established the existence of a meaningful classical limit opens
the door to the ultimate subject of our interest, the quantum
behaviour of the local gravitational degrees of freedom.
On the one hand, together with appropriate matter degrees of
freedom, they should be able to describe the quantum behaviour of
the very early universe, which may have left an imprint on the
cosmic microwave background radiation we observe today. On the
other hand, they will characterize the
fluctuations around a large-scale classical background, like that of
our present universe, leading
to quantifiable deviations from the predictions of classical relativity.

\end{document}